\documentclass{sig-alternate-05-2015}

\usepackage{epstopdf}
\usepackage{subfigure}
\usepackage{algorithmicx}
\usepackage[vlined,algoruled, linesnumbered]{algorithm2e}
\usepackage{cases}
\usepackage{color}
\usepackage{mathtools}
\DeclarePairedDelimiter{\ceil}{\lceil}{\rceil}

\usepackage{float}
\usepackage{stfloats}
\usepackage[bottom]{footmisc}
\newtheorem{Def}{Definition}
 
\begin{document}
	
%\setcopyright{acmcopyright}
%
% --- Author Metadata here ---
%\conferenceinfo{SIGMOD}{2016, USA}
%\CopyrightYear{2007} % Allows default copyright year (20XX) to be over-ridden - IF NEED BE.
%\crdata{0-12345-67-8/90/01}  % Allows default copyright data (0-89791-88-6/97/05) to be over-ridden - IF NEED BE.
% --- End of Author Metadata ---

\title{PI : a Parallel in-memory skip list based Index}
%\numberofauthors{5}
\author{
	% You can go ahead and credit any number of authors here,
	% e.g. one 'row of three' or two rows (consisting of one row of three
	% and a second row of one, two or three).
	%
	% The command \alignauthor (no curly braces needed) should
	% precede each author name, affiliation/snail-mail address and
	% e-mail address. Additionally, tag each line of
	% affiliation/address with \affaddr, and tag the
	% e-mail address with \email.
	%
	% 1st. author
%\alignauthor
%	Zhongle Xie\\
%	\email{xiezhongle@comp.nus.edu.sg}
%\alignauthor
%	Qingchao Cai\\
%	\email{caiqc@comp.nus.edu.sg}
%\alignauthor
%Beng Chin Ooi\\
%\email{ooibc@comp.nus.edu.sg}
%\alignauthor
%H.V. Jagadish\\
%\email{jag@umich.edu}
%\alignauthor
%Weng-Fai Wong\\
%\email{wongwf@comp.nus.edu.sg}
	Zhongle Xie$^{*}$, Qingchao Cai$^{*}$, H.V. Jagadish$^{+}$, Beng Chin Ooi$^{*}$, Weng-Fai Wong$^{*}$ \\
	\affaddr{$^{*}$National University of Singapore} \quad
	\affaddr{$^{+}$University of Michigan}  \\
	\email{$^{*}$\{xiezhongle, caiqc, ooibc, wongwf\}@comp.nus.edu.sg}\quad
	\email{$^{+}$jag@umich.edu}
%	\alignauthor
%	Paper ID: 618
}

\maketitle
\begin{abstract}
    Due to the coarse granularity of data accesses and the heavy use of latches, indices in 
    the B-tree family are not 
    %suitable 
    efficient
    for in-memory databases, especially in the context of today's 
    multi-core architecture.

    In this paper, we present \textbf{PI}, a \textbf{P}arallel in-memory skip list based \textbf{I}ndex
    that lends itself naturally to the parallel and concurrent environment, particularly with non-uniform memory access.  In PI,
    incoming queries are collected, and disjointly distributed among
    multiple threads for processing to avoid the use of latches.
    For each query, PI traverses the index in a Breadth-First-Search (BFS)
    manner to find the list node with the matching key, exploiting SIMD processing
    to speed up the search process.
    In order for query processing to be latch-free, PI employs a light-weight
    communication protocol that enables threads to re-distribute the query workload
    among themselves such that each list node that will be modified as a result of
    query processing will be accessed by exactly one thread.
    We conducted extensive experiments, and the results show that PI can be up to three times
    as fast as the Masstree, a state-of-the-art B-tree based index.

\end{abstract}

%\terms{}


\section{Introduction}\label{sec::intro}

DRAM has orders of magnitude higher bandwidth and lower
latency than hard disks, or even flash memory for that matter.  % WWF : flash is semiconductor
With exponentially increasing memory sizes and falling prices, 
it is now frequently possible
to accommodate the entire database and its associated indices in memory,
thereby completely eliminating the significant overheads of slow disk
accesses \cite{diaconu2013hekaton, kallman2008h, kemper2011hyper,
sikka2012efficient, tu2013speedy}.
Non-volatile memory such as phase-change memory looming on the horizon
is destined to push the envelope further.
Traditional database indices, e.g., B$^+$-trees \cite{comer1979ubiquitous}, 
that were mainly optimized for disk accesses,
are no longer suitable for in-memory databases since
they may suffer from poor cache utilization due to their hierarchical
structure, 
coarse granularity of data access and poor parallelism.

The integration of multiple cores into a single CPU chip makes many-core
computation a norm today. Ideally, an in-memory database index
should scale with the number of on-chip cores to fully
unleash the computing power.
The
B$^+$-tree index, however, is ill-suited
for such a parallel environment.
Suppose a thread is
about to modify an intermediate node in a B-tree index. 
It should first prevent other
concurrent threads from descending into the sub-tree rooted at that node
in order to guarantee correctness, thereby forcing serialization among these threads.
%{\color{red}{
Worse, if the root node is being updated, all of the
other threads cannot proceed to process queries.
%}}
%%% ooibc: one more constraint: same point of access for each query -- root 
Consequently, 
the traditional B-tree index does not provide the parallelism 
required for effective use of the concurrency provided by a many-core environment.

On the other hand, % WWF: this para a bit disjoint from the above
{\em single instruction multiple data} (SIMD) is now supported by almost all
modern processors. It enables performing the same computation, e.g.,
arithmetic operations and comparisons, on multiple data simultaneously,
holding the promise of significantly reducing the time complexity of computation.
However, to operate on indices using SIMD requires a non-trivial rethink
since SIMD operations require operands to be contiguously stored in memory.

The aforementioned issues highlight the need for a new parallelizable
in-memory index, and
motivate us to re-examine the {\em skip list}~\cite{pugh1990skip}
as a possible candidate as
the base indexing structure in place of the B$^+$-tree (or B-tree).
Skip list employs a probabilistic
model to build multiple linked lists such that each linked list consists of
nodes selected according to this model from the list at the next level.
Like B$^+$-trees,
%%%% ooibc: this statement contradicts the above weakness mentioned
%%%  
%%% the next sentence may save it -- leave it as such 1st
the search for a query key in a skip list follows
breadth-First traversal in the sense that it starts from the list
at the top level and moves forward
along this level until a node with a larger
key is encountered. Thereupon, the search moves to 
the next level, and proceeds as it did in the previous level.
However, since only one node is being touched by the search
process at any time, its predecessors at the same level
can be accessed by another search thread, which means data access in
the skip list has a finer granularity than the B$^+$-tree where an intermediate tree
node contains multiple keys, and should be accessed in its entirety.
Moreover, with a relaxation on the structure hierarchy,
a skip list can
be divided vertically into disjoint partitions,
each of which can
be individually processed on multi-core systems.
%The structure can be distributed in the same manner across multiple compute nodes. % WWF - redundant
Hence, we can expect
the skip list to be an efficient and much more suitable indexing technique for concurrent settings.

It is natural to use latches to implement concurrent accesses over a given
skip list.
Latches, however, are costly.
In fact, merely inspecting a latch 
may require it to be flushed from other caches, and is therefore costly.
Latch modification is even more expensive, 
as it will invalidate the latch
replicas located at the caches of other cores,
and force the threads
running on these cores to re-read the latch for inspection, incurring
significant bandwidth cost, 
especially in a Non-Uniform Memory Access (NUMA) architecture where accessing
memory of remote NUMA nodes can be an order of magnitude slower than
that of a local NUMA node.

In this paper, we propose a highly
parallel in-memory database index based on a latch-free skip list.
We call it \textbf{PI}, a \textbf{P}arallel in-memory skip list based \textbf{I}ndex.
%%Thus, it is easy-to-implement and performs much better in
%%concurrent scenarios, which inspired us to invent the index.
In PI,
we employ a fine-grained processing strategy to avoid using latches.
Queries are organized into batches and each batch processed simultaneously 
%%Queries are collected and distributed into different batches to process simultaneously
using multiple threads.
Given a query, PI traverses the index
in a breadth-first manner to find the corresponding list node. 
SIMD instructions are used to accelerate the search process.
To handle the case in which two threads find the same list node for some keys,
%%The resultant nodes for a batch of queries are then re-distributed among
PI adjusts the query workload among 
execution threads to ensure that each list node that will be modified as a
result of query processing is accessed by exactly one thread, thereby
eliminating the need for latches.
%%% ooibc: what do u mean? why distribute?
%%The PI indexing scheme
%%is naturally latch-free and easy to implement.
%%To ensure serializability, we
%%use delete flags and synchronization barrier among threads.

Our main contributions include:
\begin{itemize}
	\item We propose a latch-free skip list index that shows high
        performance and scalability. It serves as an
        alternative to tree-like indices for in-memory databases and suits
        the many-core concurrent environment due to its high degree of parallelism.
%        {\color{blue}
	\item We use SIMD instructions to accelerate query processing of the index.
        \item  A set of optimization techniques are employed in the proposed
            index to enhance the performance of the index in
            many-core environment.
            %%\item For the proposed index, we build an analytical model to
            %%    estimate the cost.
            %%    Our model is consistent with our experimental results.
%        }
	\item We conduct an extensive performance study on PI as well as a
        comparison study between PI and Masstree~\cite{mao2012cache}, a
        state-of-the-art index used in SILO~\cite{tu2013speedy}.
        %%{\color{red}
	    %%    We also use Yahoo! Cloud
	    %%    Serving Benchmark (YCSB)~\cite{cooper2010benchmarking} to simulate real query workload and evaluate both PI and Masstree.}
        %%% ooibc: citation for masstree and silo
        %%%
        The results show that PI is able to perform
        up to more than $3\times$ better than Masstree in terms of query throughput.
\end{itemize}

The remainder of this paper is structured as follows.
Section~\ref{sec::relate}
presents related work.
We describe the design and implementation of PI in
Section~\ref{sec::index} and \ref{sec::impl}, respectively, and develop a
mathematical model to analyze PI's performance of query processing in
Section~\ref{sec::model}.
Section~\ref{sec::eval} presents the performance study of PI.
Section~\ref{sec::conclusion} concludes this work.

\section{Related Work}\label{sec::relate}
\subsection{\text{B$^+$-tree}}

The B$^+$-tree \cite{comer1979ubiquitous} is perhaps the most widely used index in
database systems. 
%has been traditionally used as the de facto hierarchical index
%since the dawn of database systems. 
However, it has two fundamental problems
which render it inappropriate for in-memory databases 
%%{\color{blue}{compared to disk situation}}
% WWF - repetitive.
%which are becoming
%more and more popular as a result of increasing memory size and falling price.
First, its hierarchical structure leads to poor cache utilization which
in turn seriously restricts its query performance. 
Second, it does not
suit concurrent environment well
due to its coarse granularity of data access
and heavy use of latches. 
In order to solve those drawbacks, exploiting cache and removing latches 
become the core direction for implementing in-memory B$^+$-trees.	

\subsubsection{Cache Exploitation}
A better cache utilization can substantially enhance the query performance of
B$^+$-trees since reading a cache line from the cache is much faster than
from memory. 
Software prefetching is used in \cite{chen2001improving} to
hide the slow memory access. 
Rao et al.\ \cite{conf/vldb/RaoR99} presents a cache-sensitive search tree
(CSS-tree), where nodes are stored in a contiguous
memory area such that the address of each node can be arithmetically computed,
eliminating the use of child pointers in each node. 
This idea is further applied to B$^+$-trees, and the resultant structure, 
called the CSB$^+$-tree, 
%%CSB$^+$tree~\cite{rao2000making}, 
%%% ooibc: with superscript? i have changed it.
is able to
achieve cache consciousness and meanwhile support efficient update. The
relationship between the node size and cache performance of the CSB$^+$-tree is
analyzed in~\cite{hankins2003effect}. 
Masstree \cite{mao2012cache} is a trie of B$^+$-trees to efficiently handle
keys of arbitrary length. 
With all the optimizations presented in 
\cite{chen2001improving, conf/vldb/RaoR99, rao2000making} enabled, Masstree can 
achieve a high query throughput. However, its performance is still restricted 
by the locks upon which it relies to update records.
%{\color{red}
	In addition, Masstree 
is NUMA agnostic as NUMA-aware techniques have been shown to be not providing
much performance gain~\cite{mao2012cache}. As a result, expensive remote memory 
accesses are incurred during query processing.
%}

\subsubsection{Latch and Parallelizability}
There have been many works trying to improve the performance of the
B$^+$-tree by avoiding latches or enhancing parallelizability.
%%% ooibc: above two perpectives?  too embiguous
The B$^{link}$-tree~\cite{lehman1981efficient}
%%%proposed by Lehman and Yao~\cite{lehman1981efficient}, 
is an early attempt towards enhancing the
parallelizability of B$^+$-trees
 by adding to each node, except the rightmost ones,
an additional pointer
pointing to its right sibling node so that a node being modified does not
prevent itself from being read by other concurrent threads. 
However, Lomet
\cite{lomet2004simple} points out that the
deletion of nodes in B$^{link}$-trees can incur a decrease in performance and
concurrency, 
and addresses this problem through the use of additional state
information and latch coupling.
%{\color{blue}
	Braginsky and Petrank present a lock-free B$^+$-tree implemented
    with single-word CAS instructions~\cite{braginsky2012lock}.
	To achieve a dynamic structure, the nodes being split or merged will be
    frozen, but search operations are still allowed to perform against
    such frozen nodes.
%}

Due to the overhead incurred by latches, a
latch-free B$^+$-tree has also
attracted some attention. 
Sewall et al. propose a latch-free concurrent
B$^+$-Tree, named PALM~\cite{sewall2011palm}, 
which adopts bulk synchronous
parallel (BSP) model to process queries in batches. 
Queries in a batch are disjointly
distributed among threads to eliminate the synchronization among them and
thus the use of latches. 
FAST~\cite{kim2010fast} uses the same model, and
achieves twice query throughput as PALM at a cost of not being able to make
updates to the index tree. 
The Bw-tree~\cite{levandoski2013bw}, developed for
Hekaton~\cite{diaconu2013hekaton, levandoski2014indexing}, 
is another
latch-free B-tree which manages its memory layout in a 
log-structured manner and
is well-suited for flash solid state disks (SSDs) where random writes 
are costlier than sequential
ones.

\subsection{CAS Instruction and Skip List}
Compare-and-swap (CAS) instructions are atomic operations introduced 
in the concurrent environment
to ease the implementation of synchronization primitives, 
such as semaphores and mutexes.
 Herlihy proposes a sophisticated model to show that CAS instructions can be used in implementing wait-free data structures~\cite{herlihy1991wait}. 
 These instructions are not the only means to realize concurrent data structures. 
 Brown et al. also present a new set of primitive operations 
 for the same purpose~\cite{brown2013pragmatic}.

Skip lists~\cite{pugh1990skip} are considered to be an alternative to B$^+$-trees.
Compared with B$^+$-trees, a skip list has approximately the same average search
performance, but requires much less effort to implement. 
In particular, even a latch-free implementation, which is notoriously difficult
for B$^+$ trees, can be easily achieved for skip lists by using
CAS instructions~\cite{herlihy2012art}. 
Crain et al. propose new skip list algorithms~\cite{crain2013no}
to avoid contention on hot spots.
Abraham et al. combine skip lists and B-trees for efficient query processing~\cite{abraham2006skip}.
In addition, skip lists can also be integrated into distributed settings.
Aspnes and Shah present Skip Graph~\cite{aspnes2007skip} for peer-to-peer systems,
 a novel data structure leveraging skip lists to support fault tolerance.

As argued earlier,
skip lists are more parallelizable than B$^+$-trees because of the
fine-grained data access and relaxed structure hierarchy.
However, na\"{\i}ve linked list based implementation of skip lists have
poor cache utilization due to the nature of linked lists.
 In PI, we address this
problem by separating the Index Layer from the Storage Layer such that the
layout of the Index Layer is 
optimized for cache utilization and hence enables an efficient search of keys.

\subsection{Single Instruction Multiple Data}
Single Instruction Multiple Data (SIMD) processing has been extensively used in
database research to boost the performance of database operations. 
Chhugani et
al.~\cite{chhugani2008efficient} show how the performance of mergesort, 
a classical sort algorithm, can be improved, when equipped with SIMD.
Similarly, it has been shown in \cite{zhou2002implementing, balkesen2013main,
balkesen2013multi} that the SIMD-based
implementation of many database
operators, including scan, aggregation, indexing and join,
perform much better than its non-SIMD counterpart. 

Recently, tree-based in-memory indices leveraging SIMD have been proposed to
speed up query processing \cite{kim2010fast, sewall2011palm}.
FAST~\cite{kim2010fast}, a read only binary tree, can achieve an extremely high
throughput of query processing as a consequence of SIMD processing and enhanced cache
consciousness enabled by a carefully designed memory layout of
tree nodes. 
Another representative of
SIMD-enabled B$^+$-trees is PALM~\cite{sewall2011palm}, which overcomes the
FAST's limitation of not being able to support updates
at the expense of decreased query throughput.

\subsection{NUMA-awareness}
%{\color{blue}
NUMA architecture opens up opportunities for optimization in terms of
cache coherence and memory access, which can significantly hinder the
performance if not taken into 
%{\color{red}
%consideration
design
%}
~\cite{li2013numa}~\cite{blagodurov2010case}.
Since accessing the memory affiliated with remote (non-local) NUMA nodes is
substantially costlier than accessing local memory,
the major direction for
NUMA-aware optimization is to reduce the accesses of remote memory,
and meanwhile keep load balancing among NUMA nodes~\cite{psaroudakis2015scaling}.
%{\color{red}
	Many systems have been proposed with NUMA aware optimizations.
ERIS~\cite{kissinger2014eris} is an in-memory storage engine 
which employs an adaptive partitioning mechanism to realize NUMA topology
and hence reduce remote memory accesses.
%}
ATraPos~\cite{porobic2014atrapos} further avoids commutative synchronizations 
among NUMA nodes during transaction processing.

There are also many efforts devoted to optimizing database operations with NUMA-awareness. 
Albutiu et al. propose a NUMA-aware sort-merge join approach, and 
leverage prefetching to further enhance the performance~\cite{albutiu2012massively}.
Lang et al. explore how various implementation techniques for  
NUMA-aware hash join~\cite{lang2015massively}.
Li et al. study data shuffling algorithms in the context of NUMA architecture~\cite{blagodurov2010case}.
%}

\section{INDEX DESCRIPTION}\label{sec::index}

In a traditional skip list,
since nodes with
different heights are dynamically allocated, they do not reside within a
contiguous memory area.
Non-contiguous storage of nodes causes
cache misses during key search and limits
%%%incapability to  fully 
the exploitation of 
%%exploit 
SIMD processing, which requires the operands to be stored 
within a contiguous memory area.
%%% ooibc10: please check the above
%%
We shall elaborate on how PI overcomes these two limitations and meanwhile achieves
latch-free query processing.

\begin{figure*}
	\centering
	\includegraphics[width=\linewidth]{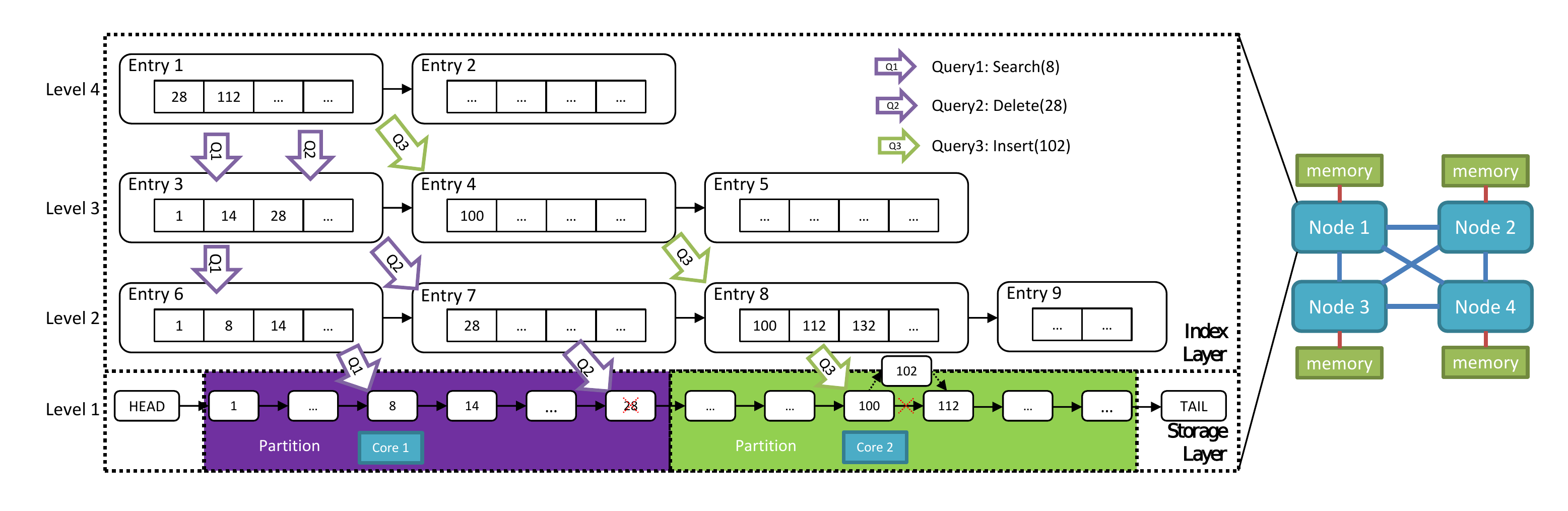}
	\caption{An instance of PI}
	\label{fig::arch}
\end{figure*}

\subsection{Structure}

Like a typical skip list, 
PI also consists of multiple levels of sorted linked lists.
The bottommost level is a list of data nodes, whose definition is given in
Definition~\ref{def::datanode}; an upper-level list is composed of the keys
randomly selected with a fixed probability from those contained in the linked
list of the next lower level. For the sake of expression,
these composing linked lists are logically separated into two layers: the
\textit{storage layer} and the \textit{index layer}. The storage layer is merely the
linked list of the bottommost level, and the index layer is made up of the
remaining linked lists.   

\begin{Def}
    A data node $\alpha$ is a triplet 
    $$(\kappa, p, \Gamma)$$
    where $\kappa$ is a
    key, $p$ is the pointer to the value associated with $\kappa$, and
    $\Gamma$ is the height of $\kappa$, representing the number of linked lists
    where key $\kappa$ is present. We say a key $\kappa \in S$ if there exists
    a data node $\alpha$ in the storage layer of the index, $S$,
    such that $\alpha.\kappa = \kappa$.
\label{def::datanode}
\end{Def}

The difference between a traditional skip list and PI lies in the index layer.
In a traditional skip list, the node of a composing linked list of the index
layer contains only one key.
In contrast, a fixed number of keys are included in a list
node of the index layer, which we shall call an \textit{entry} hereafter. The
reason for this arrangement of keys is that it enables SIMD processing, which
requires operands to be contiguously stored.
An instance of PI with four keys in an entry is shown in Figure~\ref{fig::arch},
%{\color{blue}
	where three different operations are collectively
    processed by two threads, which are represented by the two arrows
    colored purple and green, respectively.
%%	We omit some keys for the sake of simplicity.
%	}

Given an initial dataset,
the index layer can be constructed in a bottom up manner.
We only need to scan the storage layer once to fill the high-level
entries as well as the associated data structure, i.e., routing table,
which will be discussed in the next section in detail.
This construction process is $O(n)$ where $n$ is the number of records % WWF - can say something like this?
%This construction process is very fast, 
at the storage level, typically taking less than 0.2s for 16M
records when running on a 2.0 GHz CPU core.
%%% ooibc: this is data size dependent -- make a note the data size
% WWF - can you give a big-O notation for the construction?
Further, the construction process can be parallelized,
and hence can be sped up using more computing resources.

% WWF - similarly for the following, if you can give big-O running time, it would be good.
\subsection{Queries and Algorithms}\label{subquery}

PI, like most indexing structures, supports three types of queries, namely search,
insert and delete. Their detailed descriptions are as follows, and an
abstraction of them is further given in Definition~\ref{def::query}. 

\begin{itemize}
    \item\textbf{Search($\kappa$):}
        %When the optional parameter $\kappa_e$ is not given,
        %this is a point query, for which
        if there exists a data node $\alpha$ in the index 
        with $\alpha.\kappa = \kappa$, 
        $\alpha.p$ will be returned, and null otherwise. 
        %%For the range query where $\kappa_e$ is given, 
        %%the returned result set will be 
        %%$\{\alpha.p|\alpha \in S, \kappa \leq \alpha.\kappa \leq \kappa_e\}$.
	\item\textbf{Insert($\kappa$, $p$):}  
        if there exists a data node $\alpha$ in the index 
        with $\alpha.\kappa = \kappa$, 
        update the this data node by replacing
        $\alpha.p$ with $p$; 
        otherwise insert a new data node $(\kappa, p, \Gamma)$ 
        into the storage layer,
        where $\Gamma$ is drawn from a geometrical
        distribution with a specified probability parameter.	
	\item\textbf{Delete($\kappa$):}
        if there exists a data node $\alpha$ in the index 
        with $\alpha.\kappa = \kappa$, remove 
        this data node from the storage layer and return 1;
        otherwise return $null$.
\end{itemize}

\begin{Def}
    A query, denoted by $q$, is a triplet 
    $$(t, \kappa, [p])$$
    where $t$ and
    $\kappa$ are the type and key of $q$, respectively,
    and if $t$ is insert, $p$ provides the  
    pointer to the new value associated with key $\kappa$. 
\label{def::query}
\end{Def}

We now define the query set $Q$ in
Definition~\ref{def::qset}. There are two points worth mentioning in this
definition. First, the queries in a query set are in non-decreasing order of
the query key $k$, and the reason for doing so will be elaborated in
Section ~\ref{subsubsec::latchfree}.
Second, a query set $Q$ only contains point queries, and we will show
how such a query set can be constructed and leveraged to answer range queries in
Section~\ref{ssec:range}. 

\begin{Def}
    A query set $Q$ is given by
    $$ Q = \{q_i| 1\leq i\leq N\}$$
    where $N$ is the number of queries in $Q$, $q_i$ is a query defined in
    Definition~\ref{def::query}, and  
    $q_i.\kappa \leq q_j.\kappa$ \emph{iff} $i < j$.
    \label{def::qset}
\end{Def}
    
\begin{Def}
    For a query $q$, we define the corresponding interception, $I_q$, as
    the data node with the largest key among those in 
    $\{\alpha|\alpha.\Gamma > 1, \alpha.\kappa \leq q.\kappa\}$.
   %% Meanwhile, we
   %% define the interception set $\Pi$ for query set $Q$ as $\Pi = \{I_q|q\in Q\} $
\label{def::interception}
\end{Def}

\begin{algorithm}[t]
	\label{alg:query}
	\caption{{Query\ processing}}
	\small
	\SetKwInOut{Input}{Input}
	\SetKwInOut{Output}{Output}
	\SetKwFunction{redis}{redistribute}
	\SetKwFunction{partition}{partition}
	\SetKwFunction{trav}{traverse}
	\SetKwFunction{exec}{execute}
	%\SetKwFunction{getpartition}{GetPartition}
	%
	\Input{$S$, PI index \\
		$Q$, query set \\
		$t_1, ..., t_{N_T}$, $N_T$ threads\\
	}
	\Output{$R$, Result Set
	}
	$R = \emptyset$\;
	\For {$i=1 \rightarrow N_T$}
	{
		$Q_i = \partition(Q,N)$\;
	}
	/* traverse the index layer to get interceptions */ \\
	\ForEach{Thread $t_i$}
	{
		$\Pi _i = 	\trav(Q_i,S)$\;
	}
	waitTillAllDone()\;
	%%%% ooibc2: waitAllDone?  wait-till-all-done?
	/* redistribute query workload */\\
	\For {$i=1 \rightarrow N_T$}
	{
		%%$\Pi'_i = \assignwork(\Pi_1,\Pi_2,...,\Pi_{N_T})$\;
		$\redis(\Pi_i, Q_i, i)$\;
	}
	/* query execution */ \\
	\ForEach{Thread $t_i$}
	{
		$R_i = 	\exec(\Pi_i,Q_i)$\;
	}	
	waitTillAllDone()\;
	$R = \cup R_i$\;
	return $R$\;
	%}
\end{algorithm}

PI accepts a query set as input, and employs a batch technique 
%%% ooibc10: people tend to have a bit of concern when batch is mentioned
%%%   in online processing
%%%  need to somehow fix this concern
to process the queries in the input. 
Generally, batch processing may increase the latency of query processing, 
as queries may need to be buffered before being processed. 
However, since batch processing can significantly improve the 
throughput of query processing, as shown in Section~\ref{subsec:batch},
the average processing time of queries are not much affected.  % WWF - actually I find this statement not quite true - some app cares about throughput, some care about latency.

The detailed query processing of PI is given in Algorithm~\ref{alg:query}.
First, the query set $Q$ is evenly partitioned into disjoint % WWF - not divisible how?
subsets according to the number of threads, and the $i$-th subset is
allocated to thread $i$ for processing (line 3). The ordered set of queries
allocated to a thread is also a query set defined in Definition~\ref{def::qset},
and we call it a query batch in order to differentiate it from the input
query set.
Each thread traverses the index layer and generates for each query in its 
query batch an interception which is defined in
Definition~\ref{def::interception} (line 5 and 6).
After this search process, the resultant interceptions
are leveraged to adjust query batches
among execution threads such that each thread is able to  
\textit{safely} execute all the queries assigned to it after the adjustment (line
9 and 10).
Finally, each thread
individually executes the queries in its query batch (line 12 and 13).
The whole procedure is exemplified in Figure~\ref{fig::arch}, 
where three queries making up a query set are collectively processed by
three threads.
%{\color{red}
    Following the purple arrows, thread 1 traverses downwards to fetch the 
    data node with key 8 and delete the data node with key 26 in storage layer,
	and thread 2 moves along with the green arrows to insert the data node with key 102.
%	}

%%%% ooibc: query? point query only?

\subsubsection{Traversing the Index Layer}

Algorithm~\ref{alg:index} shows how the index layer is traversed to find the
interceptions for queries.
For each query key, the traversal starts from
the top level of the index layer and moves forward along this level until an
entry containing a larger key is encountered, upon which it moves on to
the next level and proceeds as it does in the previous level.
The traversal terminates when it is about to leave the bottom level of the index layer,
%%%% ooibc:???
and records the first data node that will be encountered in the storage layer as the
interception for the current query.

\begin{algorithm}[t]
	\caption{{Traversing\ the\ index\ layer}}
	\label{alg:index}
	\small
	\SetKwInOut{Input}{Input}
	\SetKwInOut{Output}{Output}
	\SetKwFunction{top}{getTopEntry}
	\SetKwFunction{isbtm}{isStorageLayer}
	\SetKwFunction{load}{\_simd\_load}
	\SetKwFunction{compare}{\_simd\_compare}
	\SetKwFunction{findnext}{findNextEntry}
	\Input{$S$, PI index \\
		$Q$, query batch\\
	}
	\Output{$\Pi$, interception set \\
	}
	$\Pi = \emptyset$\;
	\ForEach{$q \in Q$}
	{
		$e_{next} = \top(S)$\;
		$v_b = \load(q.\kappa)$\;
        \While {$\isbtm(e_{next})==\emph{false}$}
		{
			$v_a = \load(e_{next})$\;
			$mask = \compare(v_a,v_b)$\;
			/* $R_{next}$ is the routing table of $e_{next}$ */\\
			$e_{next} = \findnext(e_{next},mask,R_{next})$\;
		}
        $\Pi = \Pi \cup \{e_{next}\}$\;
	}
	return $\Pi$\;
	
\end{algorithm}

We exploit Single Instruction, Multiple Data (SIMD) processing to accelerate
the traversal.
In particular, multiple keys within an entry can be
simultaneously compared
with the query key using an \_simd\_compare instruction, which significantly
reduces the number of comparisons, and this is the main reason we put multiple keys
in each entry.
In our implementation, keys in an entry exactly occupy a
whole SIMD vector, and can be loaded into an SIMD register using a
single \_simd\_load instruction.

\begin{figure}[t]
	\centering
	\subfigure[SIMD comparison]{
	\begin{minipage}[b]{.4\linewidth}
		 \centering
		 \label{subfig::cmplogic}
		 \includegraphics[width=1.1\linewidth,height=1.1in]{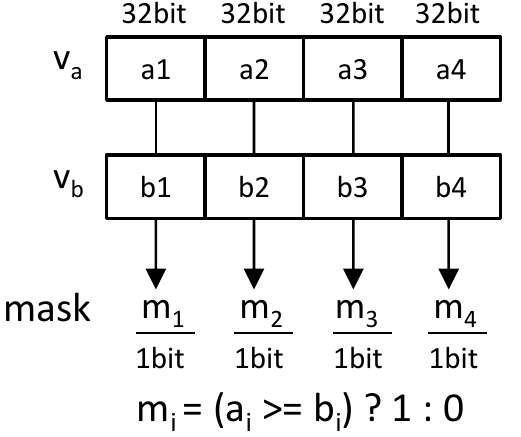} \\
		 %(a) SIMD comparison.
	\end{minipage}
	}
	\subfigure[Routing table for Entry 1]{
	\begin{minipage}[b]{.5\linewidth}
		\centering
		\label{subfig::rt2}
		\includegraphics[width=.5\linewidth]{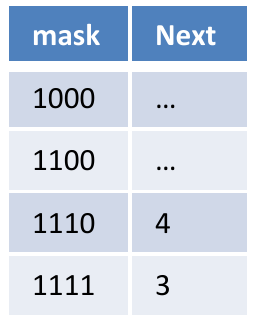} \\
		%(b) Routing table for Entry 1.
	\end{minipage}
	}
	\subfigure[Routing process for Entry 1]{
	\begin{minipage}[b]{1.\linewidth}
		\centering
		\label{subfig::rt1}
		\includegraphics[width=.9\linewidth]{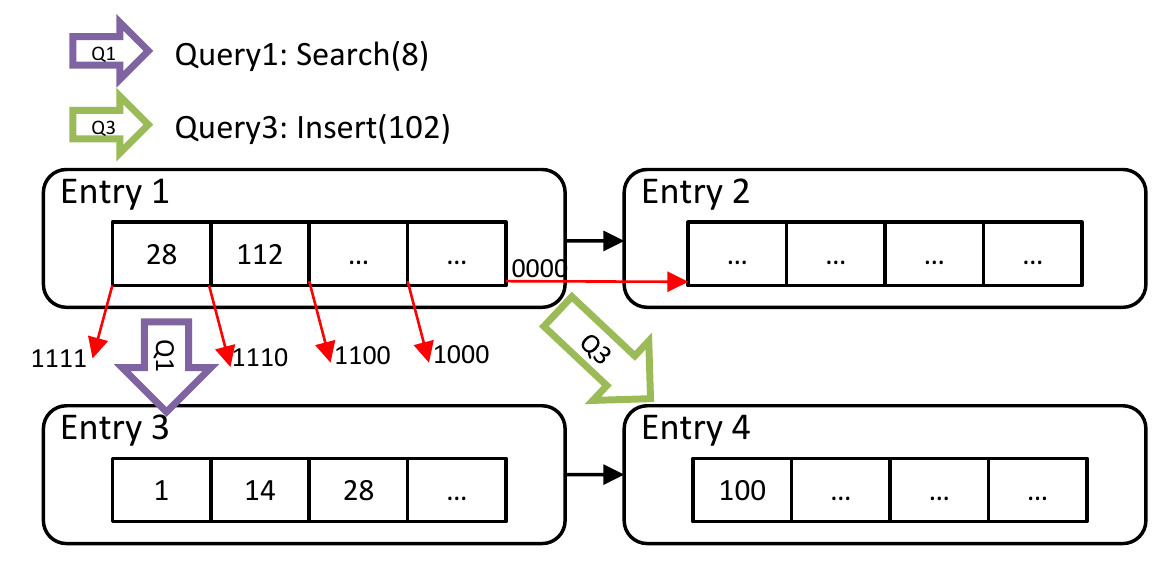} \\
		%(c) Routing process for Entry 1.
	\end{minipage}
	}
	\caption{Querying with a routing table}
	\label{fig::routetable}
\end{figure}

Since each \_simd\_compare instruction compares multiple keys in an entry,
the comparison result can be diversified,
%%% ooibc: various?
and hence an efficient
way to determine the next entry to visit for each comparison result is needed.
To this end, we
generate a routing table for each entry during the construction of PI.
%%%% ooibc: this is expensive in terms of storage space...
%%% need to show what is the cost like in the exp
For each possible result of SIMD comparison, the routing table contains
the address of the next entry to visit.
Figure~\ref{subfig::cmplogic} gives an example of SIMD comparison.
As shown
in this figure, each SIMD comparison leads to a 4-bit mask, representing
five potential results indicated by the red arrows shown in
Figure~\ref{subfig::rt1}.
This mask is then indexed into the routing
table, which is also shown in Figure~\ref{subfig::rt2}, to find the
next entry for comparison.

%%%% ooibc2: the second caption below is not readable

\subsubsection{Redistribute Query Workload}\label{sec:interception}

\begin{figure}[t]
	\centering
	\includegraphics[width=3.3in]{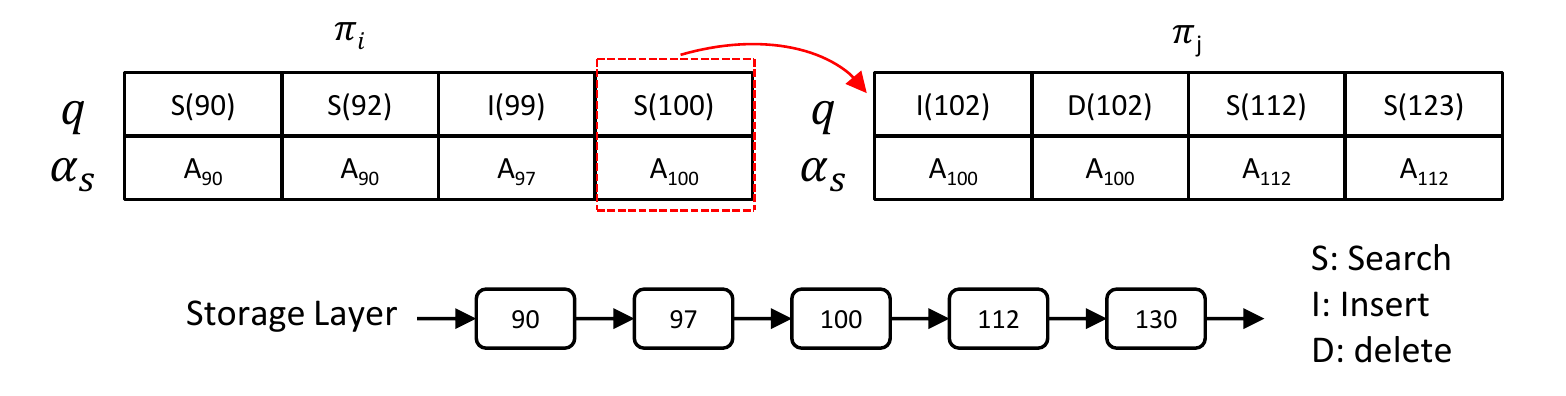}
    \caption{Interception adjustment}
	\label{fig::latchfree}
\end{figure}

Given the interception set output by Algorithm~\ref{alg:index}, a thread can
find for each allocated query $q$ the data node with the largest key that is
less than or equal to $q.\kappa$ by walking along the
storage layer, starting from $I_q$. However, it is possible that two queries
allocated to two adjacent threads have the same interception, leading to
contention between the two threads. To handle this case, we slightly adjust the
query workload among the execution threads such that each interception is
exclusively accessed by a single thread, and so are the data nodes between two
adjacent interceptions. 
To this end, each thread iterates backward over
its interception set until it finds an interception
different from the first interception of the next thread,
and hands over the queries corresponding
to the iterated interceptions to the next thread.
The details of this process are summarized in
Algorithm~\ref{alg:adjust} and exemplified in Figure~\ref{fig::latchfree}.
After the adjustment, a thread can individually execute
the allocated query workload without 
contending for data nodes with other threads.

\begin{algorithm}[!tbpn]
    \caption{Redistribute query workload}
	\label{alg:adjust}
	\small
	\SetKwInOut{Input}{Input}
	\SetKwInOut{Output}{Output}
	\SetKwFunction{recvKey}{recvKey}
	\SetKwFunction{sendKey}{sendKey}
	\SetKwFunction{recvQuery}{recvQuery}
	\SetKwFunction{sendQuery}{sendQuery}
	\SetKwFunction{recvInt}{recvInterception}
	\SetKwFunction{sendInt}{sendInterception}

	\Input{ $Q = \{q_1, q_2, ...\}$, query batch\\
        $\Pi = \{I_{q_1}, I_{q_2}, ...\}$, interception set \\
        $T_{last}$, last execution thread\\
        $T_{next}$, next execution thread\\
	}
	
	/* exchange the key of the first interception*/\\
    $\sendKey(T_{last}, I_{q_1}.\kappa)$\;
    $\kappa=\recvKey(T_{next})$\;
	
	/* wait for the adjustment from last thread */\\
	%%$\Pi' = \recvInt(T-1)$, $\Pi = \Pi' \cup \Pi$\;
    $Q' = \recvQuery(T_{last})$, $Q = Q' \cup Q$\;
    \ForEach{$q \in Q'$}
    {
        /* $I_q$ is same as $I_{q_1}$*/\\
        $\Pi = I_{q_1} \cup \Pi$
    }
	
	$Q' = \emptyset$\;
	\For {$i=|Q| \rightarrow 1$}
	{
        \uIf{$I_{q_i}.\kappa = \kappa$}
		{
            $Q' = Q' \cup \{q_i\}$, $Q = Q \setminus \{q_i\}, \Pi = \Pi \setminus \{I_{q_i}\} $\;
%%			$\Pi' = \Pi' \cup \{\Pi_i\}$, $\Pi = \Pi - \Pi_i$\;
		}
		\Else
		{
			break\;
		}
	}
	%%$\sendInt(T+1, \Pi')$\;
    $\sendQuery(T_{next}, Q')$\;
\end{algorithm}

\subsubsection{Query Execution}\label{subsubsec:execution}

The query execution process at each
thread is demonstrated in Algorithm~\ref{alg:storage}.
For each query $q$,
an execution thread iterates over the storage layer, starting from the
corresponding interception,
and executes the query against the data node with the largest key that is less
than or equal to $q.\kappa$.
If the query type is delete, we do not
remove the data node immediately from the storage layer, but merely set
a flag $F_{del}$ instead, which is necessary for latch-free query
processing, as we shall explain in Section~\ref{subsubsec::latchfree}.
For the search query, the $F_{del}$ flag of
the resultant data node will be checked to decide the validity of its
pointer.
For the update query, a new data node will be inserted into the
storage layer if the query key does not match that of the resultant data
node. Unlike a typical skip list, PI only allocates a random height
for the new node, but does not update the index layer immediately.
With more and more updates made to the storage layer, the index layer should
be updated accordingly to guarantee the performance of query processing.
Currently, a background process monitors the updates to the 
storage layer, and asynchronously rebuild the whole index layer 
when the number of updates exceeds a certain threshold. The detail is
given in Section~\ref{ssub:background_update}.
%%% ooibc10: this needs a bit of explanation, and refer to Section 5.2 for details

\begin{algorithm}[!tbpn]
	\caption{{Query execution}}
	\label{alg:storage}
	\small
	\SetKwInOut{Input}{Input}
	\SetKwInOut{Output}{Output}
	\SetKwFunction{type}{queryType}
	\SetKwFunction{get}{getNode}
	\SetKwFunction{update}{updateNode}
	\SetKwFunction{remove}{removeNode}
	\SetKwFunction{getnode}{getNode}
	\SetKwFunction{insert}{insertNode}
    \Input{ $Q = \{q_1, q_2, ...\}$, adjusted query batches\\
        $\Pi = \{I_{q_1}, I_{q_2}, ...\}$, adjusted interception set \\
	}
	\Output{ $R$, result set\\
	}
	$R=\emptyset$\;
	\For {$i=1 \rightarrow |Q|$}
	{
        /* walk along the storage layer, */ \\
        /* starting from the corresponding interception */\\
        $node = \getnode(q_i, I_{q_i})$\;
		$r_i = null$\;
		\uIf {$q_i.t == ``Search"$}
		{
			\If {$node.\kappa == q_i.\kappa\ \&\&\ !node.F_{del}$}
			{
				$r_i = node.p$\;
			}
		}
		\uElseIf {$q_i.t == ``Insert"$}
		{
			\uIf {$node.\kappa == q_i.\kappa$}
			{
				$node.p = q_i.p$\;
			}
			\Else
			{
				$node = \insert(node, q_i.\kappa, q_i.p, \Gamma)$\;
			}
			$node.F_{del} = false$\;
		}
		\Else
		{
			\If {$node.\kappa == q_i.\kappa$}
			{
				$node.F_{del} = true$\;
			}
		}
        $R = R \cup \{r_i\}$
	}
	return $R$\;
\end{algorithm}

\subsubsection{Naturally Latch-free Processing}\label{subsubsec::latchfree}

It is easy to see that query processing in PI is latch-free.
In the
traversal of the index layer, since the access to each entry is read-only,
the use of latches can be avoided at this stage.
For the adjustment of query workload,
each thread communicates with its adjacent threads via
messages and thus does not rely on latches.
In addition, each query $q$
allocated to thread $i$ after the adjustment of query workload satisfies
$I^i_{q_1}.\kappa \leq q.\kappa < I^{i+1}_{q_1}.\kappa$, where $I^i_{q_1}$ and
$I^{i+1}_{q_1}$ are the first element
in the interception sets of thread $i$ and $i+1$, respectively.
Consequently, thread $i$ can individually execute without latches all its 
queries except those which require reading $I^{i+1}_{q_1}$ or inserting a new node
directly before $I^{i+1}_{q_1}$, since the data nodes that will be accessed
during the execution of these queries will never be accessed by other threads.
The remaining queries can still be executed
without latches as the first interception of each thread will never be deleted,
as described in Section~\ref{subsubsec:execution}.

In our algorithm, a query set $Q$ is ordered mainly due to two reasons.
First, cache utilization can be improved by processing ordered queries
in Algorithm~\ref{alg:query}, since the entries and/or data nodes to be 
accessed for a query may have already been loaded into the
cache during the processing of previous queries. 
Second, a sorted query set leads to sorted
interception sets (ordered by query key), which is necessary for
interception adjustment and query execution to work as expected.

\subsubsection{Range Query}
\label{ssec:range}

%% TODO: what if the queries in a range query are distributed among two adjacent
%% threads

Range query is supported in PI.
Given a set of range queries
(a normal point query defined in Definition~\ref{def::query} is also 
a range query with the upper and lower bound of the key range
being the same),
we first sort them according to the lower bound of their key
range and then construct a query set defined in Definition~\ref{def::qset}
using these lower bounds. This query set is then distributed among a set of
threads to find the corresponding interceptions, as in the case of point
queries. The redistribution of query workload, however, is slightly different.
Denote the first element in the interception set of thread $i$, i.e., the
interception corresponding to the first query in the query batch of thread
$i$, by $I_{q_1}^i$. For each allocated query with a key range of $[\kappa_s,
\kappa_e]$, where $\kappa_e \geq I_{q_1}^{i+1}$,
thread $i$ partitions it into two queries with the key
ranges being $[\kappa_s, I_{q_1}^{i+1}.\kappa)$ and
    $[I_{q_1}^{i+1}.\kappa, \kappa_e]$, respectively,
    and hands over the second query to thread $i+1$. As in
    Algorithm~\ref{alg:adjust}, this redistribution process must be performed in
    the order of thread \textit{id} in order to handler the case where the key range of a
    range query is only covered by the union of the interception sets of three or
    more threads.
After the redistribution of query workload, each thread then executes the
allocated queries one by one, which is quite
straightforward. Starting from the corresponding interception, PI iterates over
the storage layer to find the first data node within the key range, and then
executes the query upon it and each following data node until the upper bound of the
key range is encountered. The final result of an original range query can be
acquired by combining the result of corresponding partitioned queries.

%%%%ooibc:  where do you reference this figure?????
%=======
\section{Implementation}
\label{sec::impl}

\subsection{Storage Layout}
\label{sub:storage_layout}
As we have mentioned, PI logically consists of the index layer and the
storage layer. The index layer further comprises multiple levels, each
containing the keys appearing at that level. Keys at the same level are
organized into entries to exploit SIMD processing, and each entry is associated
with a routing table to guide the traversal of the index layer. 
For better cache
utilization, entries of the same level are stored in a contiguous memory area.
The storage layer of PI is implemented as a 
%normal 
typical
linked list of data nodes to support efficient insertions.
%%% ooibc10: normal?
Since entries contained in each level of the index layer are stored compactly in a
contiguous memory area, PI cannot immediately update the index layer upon the
insertion/deletion of a data node with height $h>1$. 
Currently, we implement a simple
strategy to realize these deferred updates. Once the number of
insertions and deletions exceeds a certain threshold, the entire index layer is
rebuilt from the storage layer in a bottom-up manner. Although this strategy seems
to be time-consuming, it is highly parallelizable and the rebuilding process can
thus be shortened by using more threads. Specifically, each thread can be assigned
a disjoint portion of the storage layer and made responsible to construct the
corresponding part of the index layer. 
The final index layer can then be obtained by
simply concatenating these parts level by level.

\subsection{Parallelization and Serializability}

We focus on two kinds of parallelization in the implementation, i.e.,
data-level parallelization and scale-up parallelization. It is also
worth noting that serializability should be guaranteed in the parallel
process.

Data-level parallelization is realized through the use of SIMD
instructions during the traversal of the index layer.
As mentioned before,
in our implementation, each entry contains multiple keys, and their
comparison with the query key can be done using a single SIMD
comparison instruction, which substantially accelerates the search
process of interceptions.
Moreover, SIMD instructions can be introduced
in sorting the query set $Q$ to improve the performance.
In our
implementation, we use Intel Intrinsic Library to implement SIMD related
functions.

We exploit scale-up parallelization provided in multi-core systems by
distributing the query workload among different cores such that the execution
thread running at each core can independently process the
queries assigned to it.
Serializability issues may arise as a result of coexistence of search
and update queries in one batch,
and we completely eliminate these issues by
ensuring that only one thread takes
charge of each data node at any time.

\subsection{Optimization}
\subsubsection{NUMA-aware Optimization}

The hierarchical structure of a modern memory system, such as multiple
cache levels and NUMA (Non-Uniform Memory Access) architecture, should
be taken into consideration during query processing.
In our implementation, we organize incoming
queries into batches,
and process the queries batch by batch.
The queries
within one batch are sorted before processing.
In this manner, cache
locality can be effectively exploited,
as the search process for a query
key is likely to traverse the entries/data nodes that have just been accessed
during the search of the previous query
and thus have already been loaded into
the cache.

A NUMA architecture is commonly used to enable the
performance of a multi-core system to scale with the number of
processors/cores. In a NUMA architecture, there are multiple
interconnected NUMA nodes, each with several processors and its
own memory.
For each processor, accessing local memory residing
in the same NUMA node is much faster than accessing remote
memory of other NUMA nodes. It is thus extremely important for
a system running over a NUMA architecture to reduce or even
eliminate remote memory access for better
performance.

%PI employs NUMA-aware techniques to fully utilize the computing power in a NUMA architecture, while incurring as little
%remote memory access as possible. % WWF - this just repeat what was said above
%	Specifically,
	PI evenly distributes data nodes among available NUMA nodes
	such that the key ranges corresponding to the data nodes
	allocated to each NUMA node are disjoint.
    One or more threads will then be spawned at each NUMA node
    to build the index from the data nodes of the same NUMA node,
    during which only local memory accesses are incurred.
As a result,
for each NUMA node, there is a separate
index, which can be used to independently answer queries falling
in the range of its key set.
Each incoming query will be routed to the corresponding NUMA node, 
and get processed by the threads spawned in that node.
In this way, there is no remote memory access during
query processing, which will translate into significantly enhanced query
throughput, as shown in the experiments.
Moreover, the indices at different NUMA
nodes also collectively improve the degree of parallelism since they can be used
to independently answer queries.
%{\color{red}
	This idea of parallelism is aptly illustrated in Figure~\ref{fig::arch}, where two threads are spawned in the first NUMA node and the other three nodes can have multiple threads running in parallel as well.
%	An example is shown in Figure~\ref{fig::arch},
%    There are four NUMA nodes shown at the right side. The 
%    index built in the first NUMA node is illustrated at the left side, and
%    two threads are spawned in the same node to process Query 1-3.
%}
%%%ooibc2:
%%% talk about each separate index can be searched/accessed independently
%%% and so, it improves the degree of parallelism
%%%

\subsubsection{Load Balancing}
\label{subsec:load}

It can be inferred from Algorithm~\ref{alg:adjust} that PI
ensures the data nodes between two adjacent interceptions are handled by
the same thread.
As long as the queried keys of a batch are
not limited to a very short range, PI is able to distribute the query
workload evenly among multiple threads (within a NUMA node) due to the fact that queries of a batch
are sorted and the way we partition the queries.
However, it is more
likely that the load among multiple NUMA nodes is unbalanced, which
occurs when most incoming queries should be answered by the data nodes
located at a single NUMA node.
To address this problem, we use a
self-adjusted threading mechanism.
In particular, when a query batch
has been partitioned and each partitioned sub-query has been assigned to
the corresponding NUMA node,
we spawn threads in each NUMA node to
process its allocated queries such that the number of threads in each
NUMA node is proportional to the number of queries assigned to it. 
Consequently,
a NUMA node that has been allocated more queries will also 
spawn more threads to process queries.
An example and its further explanation on the threading mechanism are given in
Section~\ref{sssub:threading}.
\begin{figure}[tbp]
	\centering
	\subfigure[Uniform query workload] {
		\includegraphics[width=.45\textwidth]{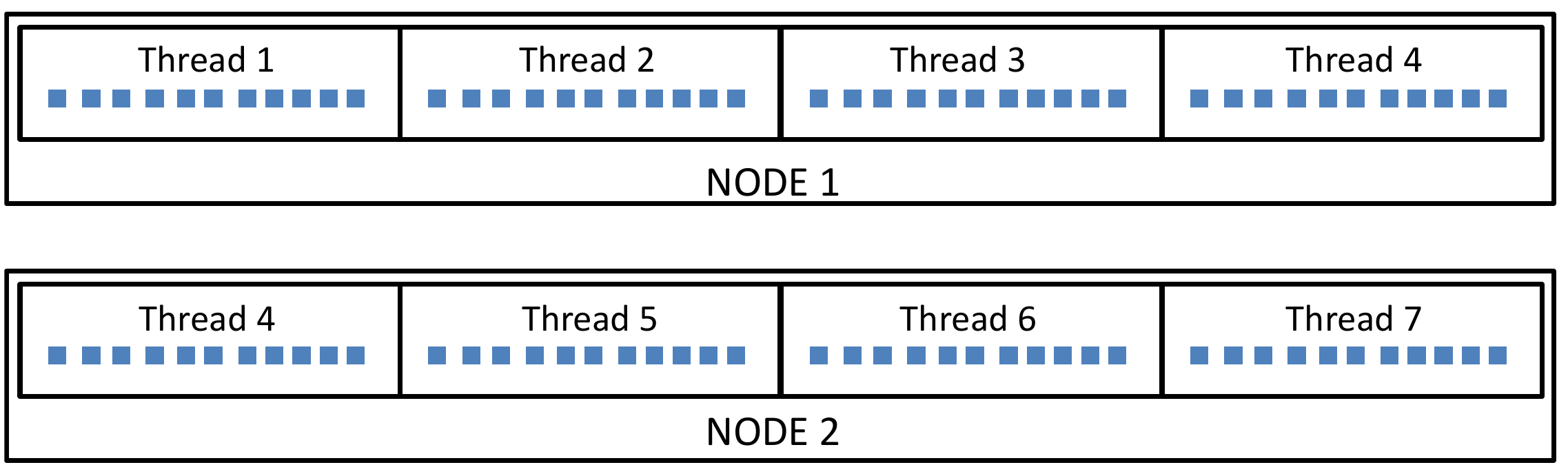}
		\label{fig:threadnormal}
	}
	\subfigure[Skewed query workload] {
		\includegraphics[width=.45\textwidth]{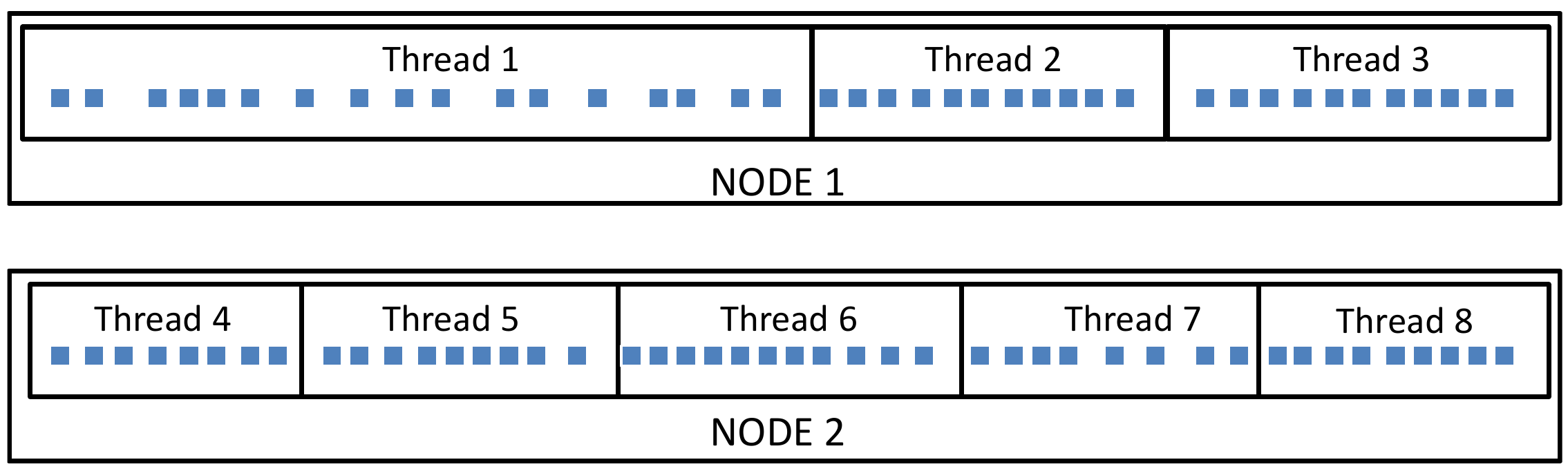}
		\label{fig:threadskew}
	}
	\caption{Self-adjusted threading}
	\label{fig:threadmanage}
\end{figure}
%%% ooibc2:
%%%  so? what is the logic?
%%%  This is to handle skewed query workload.
\subsubsection{Self-adjusted threading}
\label{sssub:threading}

In this section, we shall elaborate on 
our self-adjusted threading mechanism, which allocates threads among NUMA nodes 
for query processing 
such that query performance can be maximized within a given budget of computing
resource. As mentioned in Section~\ref{subsec:load}, if there are several NUMA nodes
available, PI will allocate the data nodes among them, build a separate
index in each NUMA node from its allocated data nodes,
and route arriving queries to the NUMA node with matching keys in order not to
incur expensive remote memory accesses.  
Given the query workload at each NUMA node, our
mechanism dynamically allocates execution threads among NUMA nodes such that 
the number of threads running on each NUMA node is proportional to its query
workload, i.e., the number of queries routed to it. In the cases where 
a NUMA node has used up its hardware threads, our mechanism will offload
part of its query workload to other NUMA nodes with available computing resource. 
\subsubsection{Group Query Processing and Prefetching}
\label{ssub:prefetch_and_group_execution}
Since entries at the same level of the index are stored in
a contiguous memory area, it is thus more convenient (due to fewer translation lookaside buffer misses)
to fetch
entries of the same level than to fetch entries locating at
different levels. 
In order to realize this location proximity,
instead of processing queries one by one,
PI organizes the queries of a batch into multiple
query groups, and processes all queries in a group simultaneously.
At each level of the index layer, PI traverses
along this level to find the first entry at the next level to compare with 
%for each query in the group being processed,  % WWF - this sentence don't make sense
and then moves downward to the next lower level
to repeat this process.  % WWF - not sure what you mean here.

Moreover, this way of group query processing
naturally calls for the use of prefetching.
When the entry to be compared with at the next level
is located for a query, PI issues a prefetch instruction
for this entry before turning to the next query. Therefore,
the requested entries at the next level may have
already been loaded into L1 cache before the comparison % WWF - Cannot say for sure L1. Prefetch implementation is machine dependent.
between them and the queried keys, thereby overlapping the
slow memory latency.

\subsubsection{Background Updating}
\label{ssub:background_update}
%%We keep two copies of index Layer in PI, one for query processing and the other for index Layer updating. 
We use a daemon thread running in background to update index layer. 
If the number of update operations meets a pre-defined threshold, 
the daemon thread will start rebuilding the index layer.
The rebuilding of the index layer is fairly straightforward. 
The daemon thread traverses the storage layer,
put the key of each valid data node with height $ > 1$ encountered
into an array sequentially,
and updates the associated route table with the address of this data node.
The new index layer will be put into use after all running threads 
complete the processing of current query batch, 
and meanwhile the old index layer will be discarded.
\\
\\

\section{Performance Modeling}
\label{sec::model}

In this section,
we develop a performance model for query processing
of PI.
The symbols used in our
analysis are summarized in 
Table~\ref{tab:analysis}.

\begin{table}
	\centering
	\caption{Notations for Analysis}\label{tab:analysis}
	\begin{tabular}{|l|l|}
		\hline
		Symbol & Description \\
		\hline
        {$H$} & index height\\
		\hline
        {$P$} & probability parameter\\
		\hline
        {$M$} & number of keys contained in an entry\\
		\hline
        {$L$} & memory access latency\\
		\hline
        {$N$} & number of the initial data nodes \\
        \hline
        {$R$} & ratio of insert queries \\
        \hline
        {$S_{e}$} & entry size \\
        \hline
        {$S_{n}$} & data node size \\
        \hline
        {$S_{l}$} & size of a cache line \\
        \hline
        {$S_{c}$} & size of the last level cache \\
        \hline
        {$T_{c}$} & time to read a cache line \\ 
                  & from the last level cache \\
		\hline
	\end{tabular}
\end{table}

\subsection{Key Search}
\label{sub:key_search}

The key search process is to locate the data node with matching key at Storage
Layer. The time spent in this process is dominated by the number of
entries and data nodes that need to be read for key comparison.

Given the number of initial data nodes,
$N$, the height of PI, $H$, is
about $\ceil{-\log_PN}$.
At each level of PI,
the number of keys between two adjacent ones that
also appear at a higher level follows a geometric distribution, and has
an expected value of $1/P$.
%{\color{blue}
Therefore, the average number of entries
needed to compare with at each level of the index layer is approximately
$\ceil{\frac{1+P}{2PM}}$, where $\frac{1+P}{2P}$ is the average number of keys
that need to be compared with. The number of cache lines that need to be read at each level
is thus $\ceil{\frac{S_{e}}{S_{l}}}\ceil{\frac{1+P}{2PM}}$.
Consequently, the total number of cache lines
%loaded from the last level cache or memory caused by
during the traversal of the index layer is
$(H-1)\ceil{\frac{S_{e}}{S_{l}}}\ceil{\frac{1+P}{2PM}}$.
%}
This is however a slight over-estimate,
since the top levels of the index layer may already have been read into L1 cache.

When the interception has been located in the storage layer for a given query,
the search process proceeds by iterating over the data nodes from
the one contained in the interception until the node with matching key
is encountered.
The number of data nodes scanned during this phase is
about half of the number of data nodes contained between two adjacent
ones with their key appearing at the index layer, which is given by
$\ceil{\frac{1+P}{2PM}}$, and the number of cache lines read during this stage is
$\ceil{\frac{S_n}{S_l}}\ceil{\frac{1+P}{2PM}}$.

The number of data nodes read during the scanning of the storage layer is not
affected by the delete queries because of the PI query processing strategy.  
However, insert queries do impact this number.
Assuming insert queries are uniformly distributed among the storage layer, and the
aggregate number of insert and delete queries does not exceed the threshold
upon which the rebuilding of the index layer will be triggered. The number of data
nodes during key search process will become $\frac{(1+iR/N)(1+P)}{2P}$, where $i$ is the
number of queries that have been processed, and $R$ is the ratio of insert
queries among the processed $i$ queries.

In our implementation, the probability of key elevation, $P$, is 0.25
as suggested in~\cite{pugh1990skip},
%%%
%%% ooibc10: where do you get the number 0.25?
%%% need to justify why 0.25, but not some other number
%%%
and each entry contains 4 keys, each being a 32-bit float number.
Therefore, the size of each entry, $S_e$, is 4*(4+8)=48
bytes, where the additional 8 bytes for each key is required by
route table to determine the next entry to compare. Each data node
occupies 20 bytes: 4 bytes are used for key, and the other 16 bytes
compose two pointers, one pointing to the value, e.g., a tuple in a
database table, and the other for the next data node. Consider an
instance of PI with 512K keys.  Its size can be computed by
$20 * 512K + 1/3 * 48 * 512K = 18M$,
 and the whole index can be kept in
the last level cache of most servers (e.g. the one we use for the experiments).
Therefore,
the processing of each search/delete query fetches about 12 cache lines, 
%{\color{blue}
9 for the index layer and 3 for the storage layer.
The total time cost is thus $12T_c$, where $T_c$ is the time to read a cache line from
the last level cache.
%}
%which will be confirmed in our experimental study.
%%%% ooibc: use a standard number multiplied by 12 (with a reference)
%%% and then
%%%  say,
%%% this is confirmed in our experimental study.

\subsection{Rebuilding the Index Layer}
\label{sub:rebuilding_index_layer}
For the sake of simplicity, we only focus on the indices with a lot of data
nodes, in which case the data nodes are unlikely to be cached, and hence
require to be read from memory during the rebuilding of the index layer.
In addition, a data node occupies 48 bytes, as mentioned in last section, and
hence can be fetched within a single memory access, which normally brings a 
64-byte cache line into the cache. Therefore, for an index with $N$ data nodes,
scanning the storage layer costs a time of $NL$. In addition, there are
$NP/(1-P)$ entries and routing tables that need to be written back into
memory, which costs another $2NP/(1-P)$ memory accesses. Therefore, the total
time of rebuilding the index layer can be approached by $(1+P)NL/(1-P)$.
However, with more threads participating in the rebuilding process, this time 
can be reduced almost linearly before bounded by memory bandwidth.
\begin{figure}[tb]
	\centering
	\includegraphics[width=1.0\linewidth]{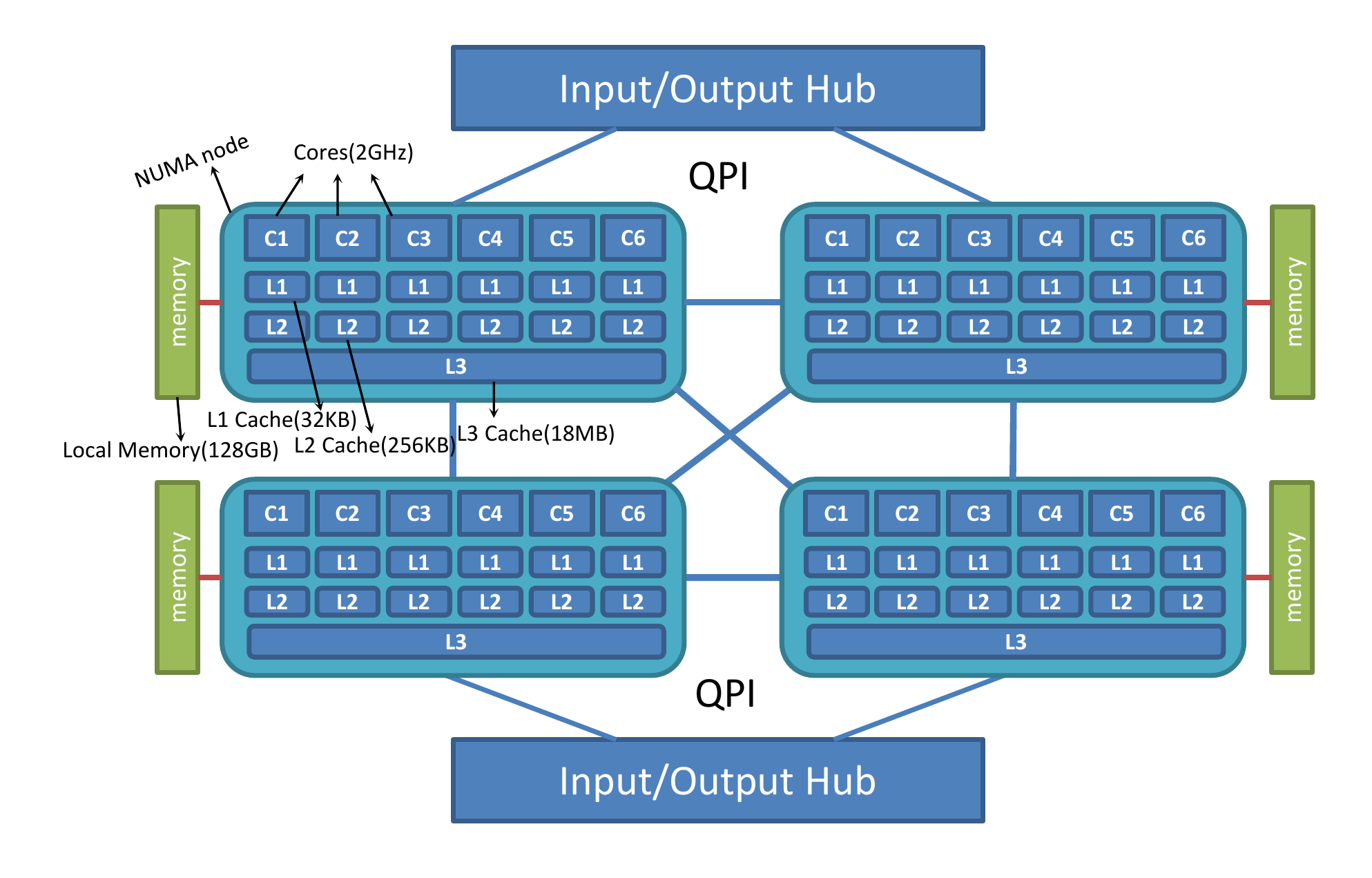}
	\caption{CPU architecture for the experiments}
	\label{fig:cpu}
\end{figure}

\section{Performance Evaluation}
\label{sec::eval}

We evaluate the performance of PI on a platform with 512 GB of memory evenly
distributed among four NUMA nodes. Each NUMA node is
equipped with an Intel Xeon 7540 processor, which supports 128-bit wide SIMD
processing, and 
has a L3 cache of 18MB and six on-chip cores, each running at 2 GHz. 
%{\color{blue}
The CPU architecture is described in Figure~\ref{fig:cpu},
    where QPI stands for Intel QuickPath Interconnect.
%    }
The operating system installed in the
experimental platform is Ubuntu 12.04 with kernel version 3.8.0-37.

\begin{table}
	\centering
	\caption{Parameter table for experiments}\label{tab:para}
	\begin{tabular}{|l|l|}
		\hline
		Parameter & Value \\
		\hline
		Dataset size(M) & 2, 4, 8, \underline{16}, 32, 64, 128, 256\\
		\hline
		Batch size & 2048, 4096, \underline{8192}, 16384, 32768\\
		\hline
		Number of Threads & 1, 2, 4, \underline{8}, 16, 32\\
		\hline
		Write Ratio(\%) & 0, 20, 40, 60, 80, 100\\
		\hline
		Zipfian parameter {$\theta$} & \underline{0}, 0.5, 0.9 \\
		\hline
	\end{tabular}
\end{table}
The performance of PI is extensively evaluated from various perspectives.
First, we show the adaptivity of PI's performance of query processing by varying
the size of the dataset and batch, and then adjust the number of execution
threads to investigate the scalability of PI. 
Afterwards, we study how PI performs in the presence of mixed and skewed query
workloads, and finally examine PI's performance with respect to range query. 
For comparison, the result of Masstree~\cite{mao2012cache} under the
same experiment setting is also given, whenever possible. 
%{\color{blue}
	We choose Masstree as our baseline mainly due to its high performance
    and maturity which is evidenced by its adoption in a widely recognized system, 
    %{\color{red}
    namely
SILO~\cite{tu2013speedy}.
%}
The Masstree code we use is retrieved from github~\cite{masstreesource}, and its
returned results are consistent with (or even better than) those presented in the
original paper~\cite{mao2012cache}, as shown in the following sections. %}

If not otherwise specified,
we use the following default settings for the experiments.
The key length is four bytes.
There are eight execution threads running on the four NUMA nodes, and the number
of threads running on each node is proportional to the query workload for this
node, as mentioned in Section~\ref{sub:storage_layout}.
Three datasets, each with a different number of keys, are used.
The small and medium datasets have 2M and 16M keys, respectively,
and the large dataset has 128M keys. 
The index built from the dataset
are evenly distributed among the four NUMA nodes.
Each NUMA node holds a separate index accounting
for approximately 1/4 of the
keys in the dataset.
%%{\color{blue}
	All the parameters for the experiments are summarized in
    Table~\ref{tab:para}, where the default value is underlined when applicable.
%%	}

%{\color{blue}
	The query workload is generated from \textit{Yahoo! Cloud
    Serving Benchmark} (YCSB)~\cite{cooper2010benchmarking}, and
the keys queried in the workload follow a zipfian distribution
with parameter $\theta = 0$, i.e., a uniform distribution. 
%}
A query batch, i.e., the set of queries allocated to a thread after 
partitioning in Algorithm~\ref{alg:query}, contains 8192 queries, 
and a discussion on how to tune this parameter is given in
Section~\ref{subsec:batch}.
The whole index layer is asynchronously rebuilt
after a fixed number
(15\% of the original dataset size) of data nodes have been
inserted/deleted into/from the index.

\subsection{Dataset Size}
\label{subsec:dataset}

\begin{figure}[tb]
	\centering
	\includegraphics[width=0.8\linewidth]{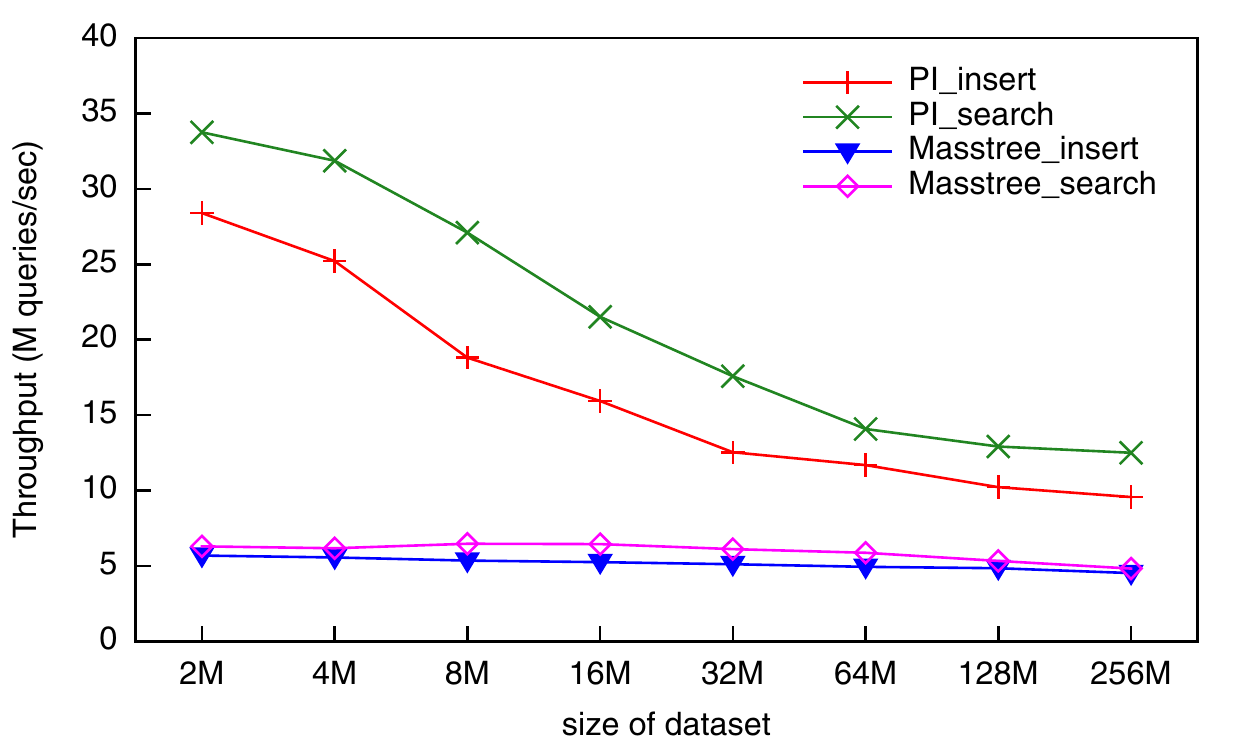}
	\caption{Query throughput vs dataset size}
	\label{fig:dataset}
\end{figure}
Figure~\ref{fig:dataset} shows the processing throughput of PI and Masstree
for search and insert queries.
For this experiment, we
vary the number of keys in the dataset from 2M to 256M,
and examine the query throughput for each dataset size.
The whole index and query workload is evenly distributed 
among the four NUMA nodes, and there are two threads running
over each NUMA node to process queries.

From Figure~\ref{fig:dataset},
one can see that the throughput for both search and insert
 experiences a moderate decrease as the dataset size increases from
2M to 64M,
and then becomes relatively stable for larger dataset sizes.
This variation
trend in the throughput is natural.
For the dataset with 2M keys,
as we have explained in Section~\ref{sec::model},
the entire index can be
%%% ooibc: total-->whole/entire
accommodated in the last level caches of the four NUMA nodes,
and the throughput is hence mainly determined by the latency 
to fetch the entries and data nodes from the cache.
As the dataset size increases,
more and more entries and data nodes are no longer able to reside in the cache
and hence can only be accessed from memory,
resulting in higher latency and lower query throughput.  

%{\color{blue}
The throughput of insert queries of PI is not as high
as that of search queries. The reason is two-fold.
First, as insert queries are processed, more and more
data nodes are inserted into the storage layer of the index, 
resulting in an increase in the time to iterate over the 
storage layer. Second, the creation of data nodes leads to
the eviction of entries and data nodes from the cache, 
and a reduced cache hit rate. This also explains why the performance gap
between insert and search queries gradually shrinks with the size of dataset. 
%}

As can be observed from Figure~\ref{fig:dataset},
the throughput of both search and
insert queries in Masstree is much less than that of PI.
In particular, PI is
able to perform 34M search queries or 29M insert queries in one second
when the index can be
accommodated in the cache,
which are respectively five and four times more than 
the corresponding throughput 
Masstree achieves for the same dataset size.
%%% ooibc: make very sure that this is correct
%%% and make sure there is no bug
%%%  the results and the index are correct
For larger datasets, PI can
still consistently perform at least 1.5x and 1x
better than Masstree in terms of the
search throughput and insert throughout, respectively.

\subsection{Batch Size}
\label{subsec:batch}

We now examine the effect of the size of query batches, 
%, i.e, the number of % WWF - just said above
%queries allocated to each thread after the partition of a query set, 
on the 
throughput of PI.
For this experiment, the three default datasets,
i.e., the small, medium and large datasets
with 2M, 16M and 128M keys, respectively, are used.
%Jag commented out.  DonÕt see the reason to say this here.  Not referenced below.
%The reason for such
%selection is that for the small and large datasets,
%the time spent in query processing is respectively dominated by cache and memory
%accesses, while for medium dataset, the accesses of cache and memory both have a
%comparable impact.

\begin{figure}[tb]
	\centering
	\includegraphics[width=0.8\linewidth]{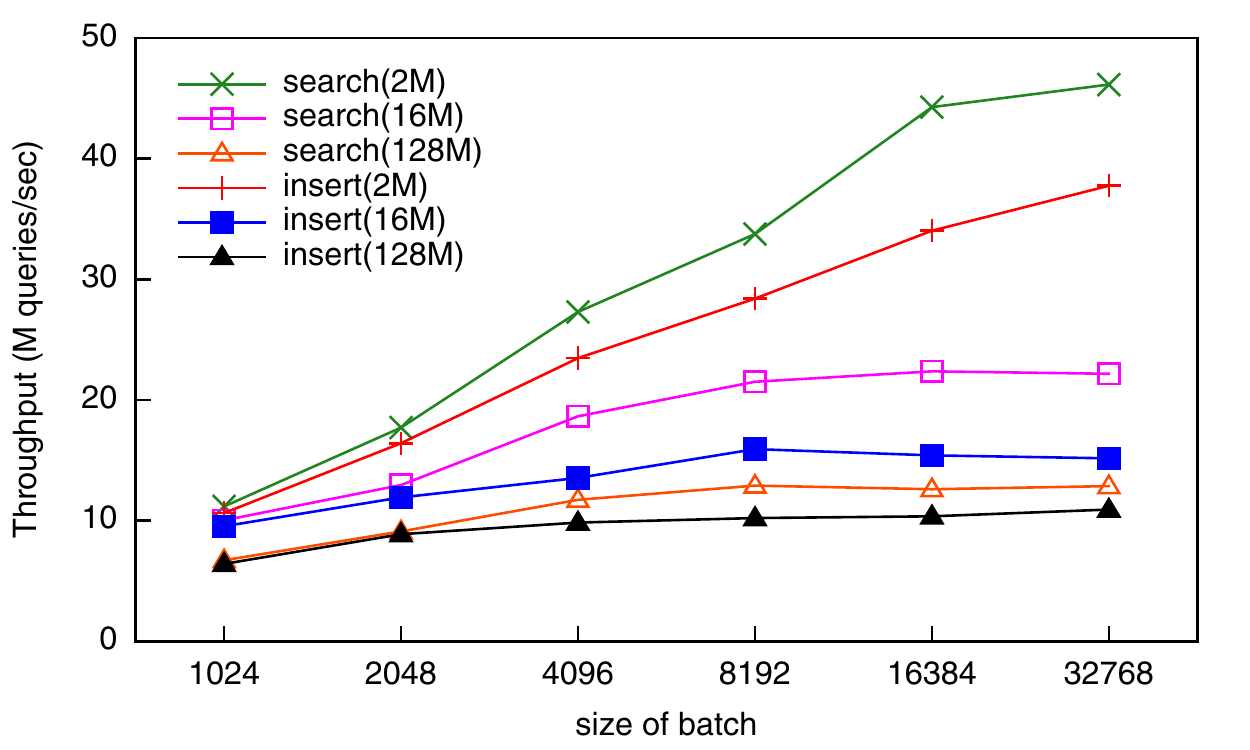}
	\caption{Query throughput vs batch size}
	\label{fig:batch}
\end{figure}

Figure~\ref{fig:batch} shows the result of query
throughput with respect to query batch size for the three datasets.
It can be seen that the size of query
batches indeed affects query throughput.
In particular, as the size of
query batch increases,
the throughput first undergoes a moderate increase.
This is due to the fact
that the queries contained in a
batch are sorted based on the key value, and
a larger batch size implies a better utilization
of CPU caches.
In addition, there is an interception adjustment
stage between key search and query execution
in the processing of each
query batch, whose
cost only depends on the number of running threads, and thus is similar
across different batch sizes.
Consequently, with more queries in a
single batch, the number of interception adjustments can be reduced,
which in turn translates into an increase in query throughput.

Figure~\ref{fig:batch} also
demonstrates that the effect of batch size exerting on query throughput is
more significant for smaller datasets than for larger datasets.
The reasons are as follows.
For query batches of the same size, the
processing time increases with the size of dataset,
as we have already
shown in Figure~\ref{fig:dataset}.
%{\color{blue}
Therefore, for smaller datasets, the
additional time spent in interception adjustment and warming up the cache, 
which is similar across the three datasets,
plays a more important role than
for larger datasets.
%}
As a result, smaller datasets benefit more from the increase in query batch size
than larger datasets do.
%%%% ooibc3: the sentence is too long and complex
%%%% shorten it by breaking it up!
It can be seen from Figure~\ref{fig:batch} that PI performs reasonably well
under the default setting of 8192 for batch size, but there still remains space
of performance improvement for small datasets. Hence, there is no
one-size-fits-all optimal setting for batch size, and we leave behind
the automatic determination of optimal batch size as future work.

\subsection{Scalability}
\begin{figure*}[tbp]
	\centering
	\subfigure[dataset size = 2M] {
		\includegraphics[width=.30\textwidth]{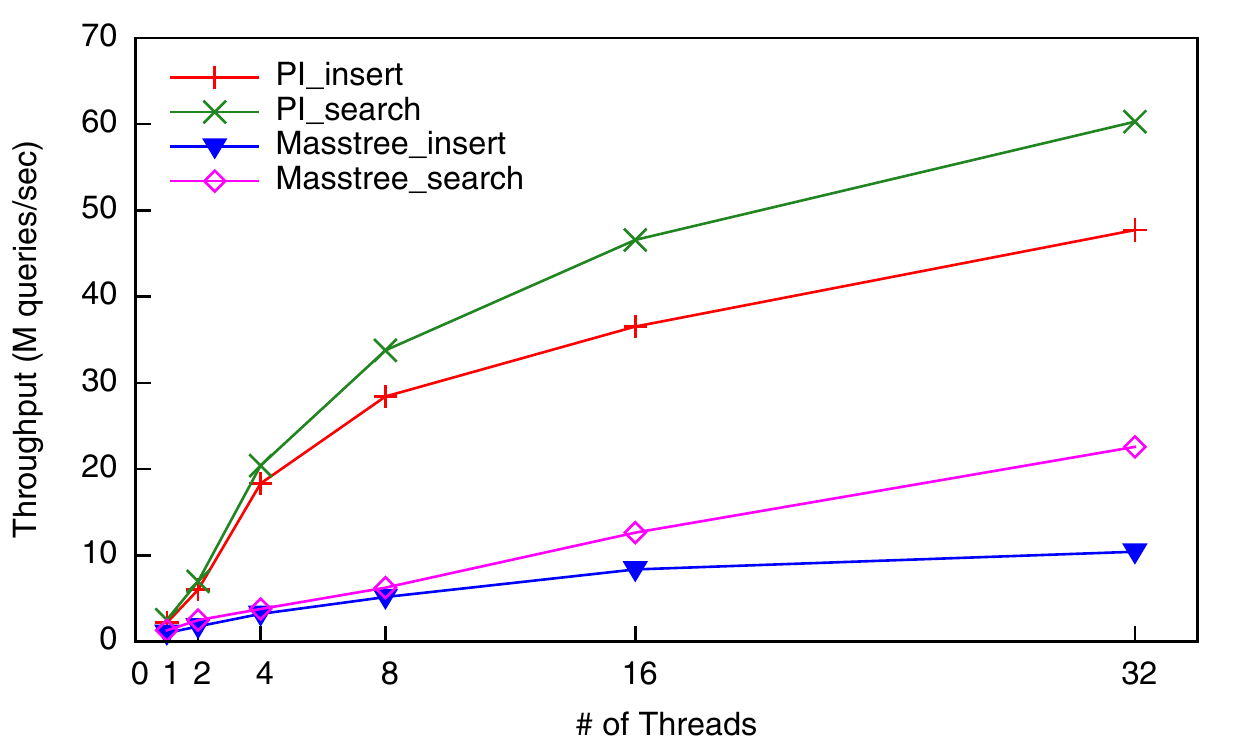}
		\label{fig:thread:2m}
	}
	\subfigure[dataset size = 16M] {
		\includegraphics[width=.30\textwidth]{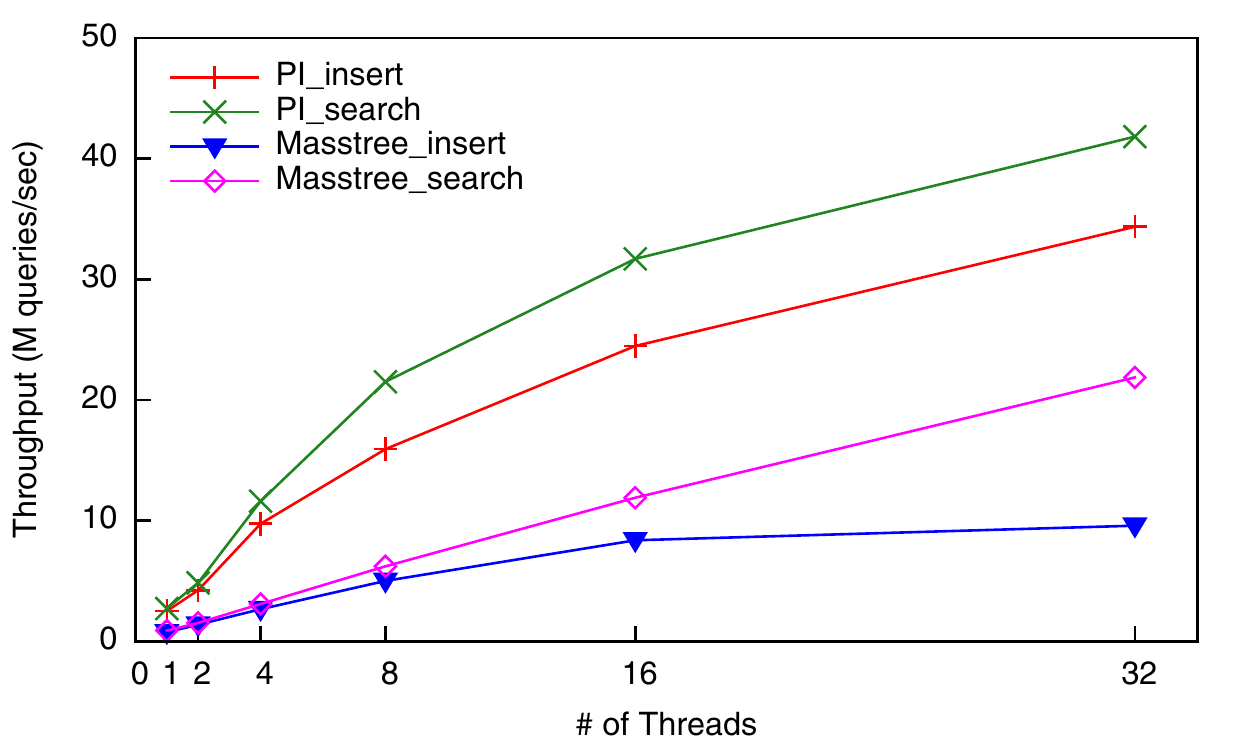}
		\label{fig:thread:16m}
	}
	\subfigure[dataset size = 128M] {
		\includegraphics[width=.30\textwidth]{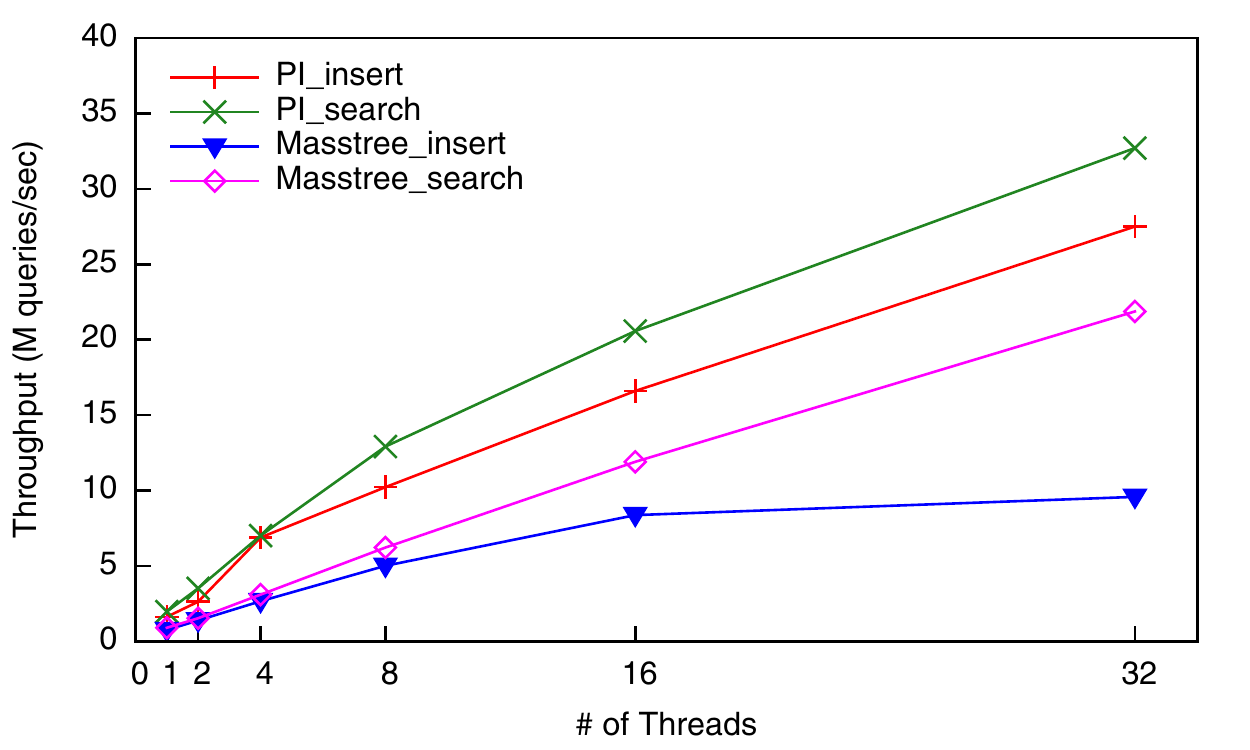}
		\label{fig:thread:128m}
	}
	\caption{Query throughput vs thread number}
	\label{fig:thread}
\end{figure*}

Figure~\ref{fig:thread} 
shows how the query throughput of PI and of Masstree varies
with the number of execution threads.
The threads and the index are both evenly distributed among the
four NUMA nodes.
For the case in which there are $n < 4$ execution threads, threads and
the whole index are evenly distributed among the same number of NUMA nodes which
are randomly selected from the available four.
We use the three datasets, i.e., 2M, 16M and 256M respectively, for this experiment.

It can be seen from Figure~\ref{fig:thread} that
apparently, both PI and Masstree can get their query throughput to 
increase significantly with more computing resources, but there exists a substantial gap in 
the rate of improvement between PI and Masstree, especially in the cases with a
small number of threads.
In PI, when the number of threads
changes from 1 to 4, the throughput undergoes a super-linear increase. 
%{\color{blue}
    This is because when the number of threads is no larger than 4, the whole index is
    evenly distributed among the same number of NUMA nodes. Consequently,
    the index size, i.e., the number of entries and data nodes, halves when the number
of threads doubles.
%} 
With a smaller index size, cache can be more effectively
utilized, leading to an increased single-thread throughout, and hence a
super-linear increase in aggregate throughout. Since smaller indices are more cache
sensitive than larger ones, they benefit more from the increase of the number of
threads in terms of the throughput of query processing, as can be observed from
Figure~\ref{fig:thread}.

When the number of threads
continues to increase, the increase rate 
in query throughput of PI gradually slows down,
but is still always better than or equal to that of Masstree. 
This flattening can
be attributed to the adjustment of interceptions which take place when there are
more than one thread servicing the queries in a NUMA node. And since the cost of
interception adjustment only depends on the number of threads, it is same across the
indices with different sizes, and hence accounts for a larger fraction of query
processing time for smaller indices. This also explains why query
throughput of smaller indices drops faster than that of larger ones. % WWF: honestly, I don't quite get what is said in this paragraph.

\subsection{Mixed Query Workload}

\begin{figure}[tbp]
	\centering
	\includegraphics[width=.9\linewidth]{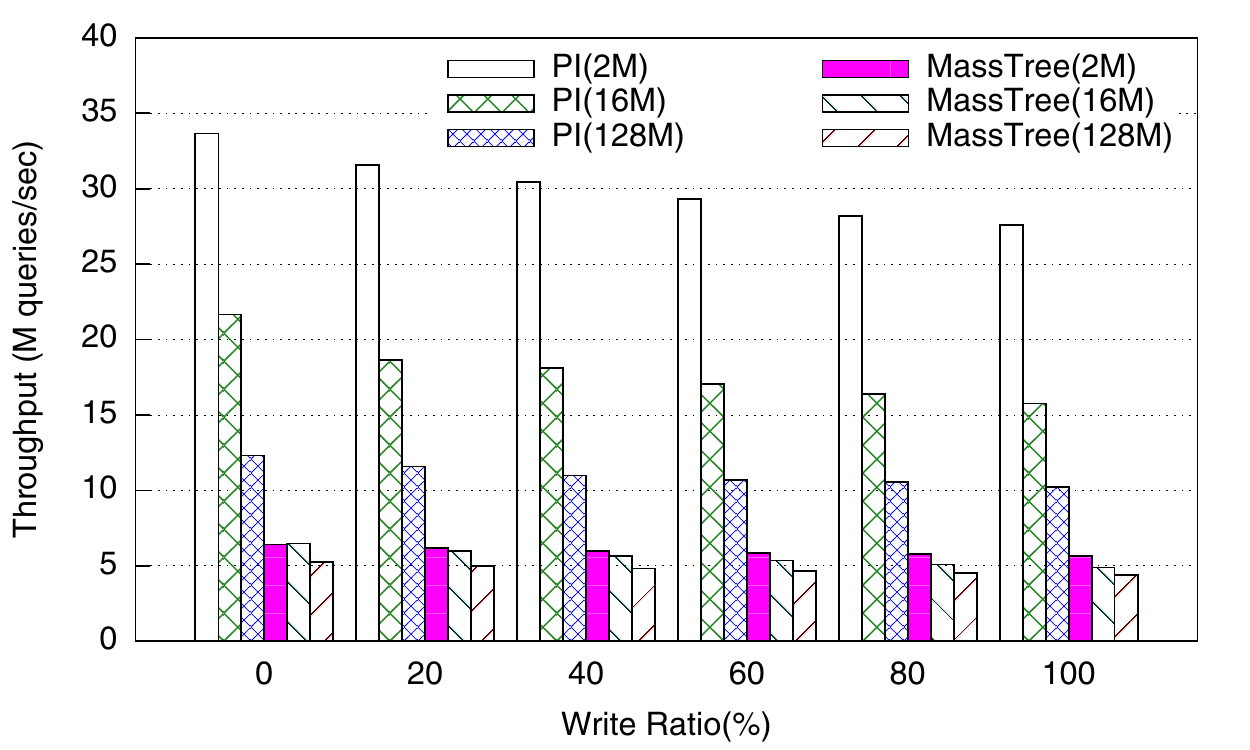}
	\caption{Query throughput of mixed workloads}
	\label{fig:mix}
\end{figure}
In order to thoroughly profile the performance of PI,
we study how it
behaves in the presence of query workload consisting of 
various types of queries.
As before, we conduct this
experiment over three default datasets 
%{\color{blue}
(2M, 16M and 128M).
%}.
The query workload consists of keys following a uniform distribution,
and the entire index layer is rebuilt
once a specified number (15\%
of dataset size) of data nodes have been inserted into the index.

Figure~\ref{fig:mix} shows the result when the ratio of insert queries
increases from 0\% to 100\%. For each ratio, the number of queries issued
to the index is such that the index layer is rebuilt for the
same number as for other ratios.
As shown in Figure~\ref{fig:mix}, with the increase of updates in the query
workload, the throughput of PI undergoes a slight decrease as a result
of more data nodes in storage layer being traversed, demonstrating 
PI's capability in processing uniform query workload. 
%{\color{blue} 
	Masstree also experiences a similar variance trend 
    in the query throughput.
%In the following section, we shall investigate how PI can handle skewed queries.
%}

\subsection{Resistance to Skewness}
\label{ssub:skew}
\begin{figure*}[tb]
	\centering
	\subfigure[$\theta=0.5$] {
		\includegraphics[width=.40\textwidth]{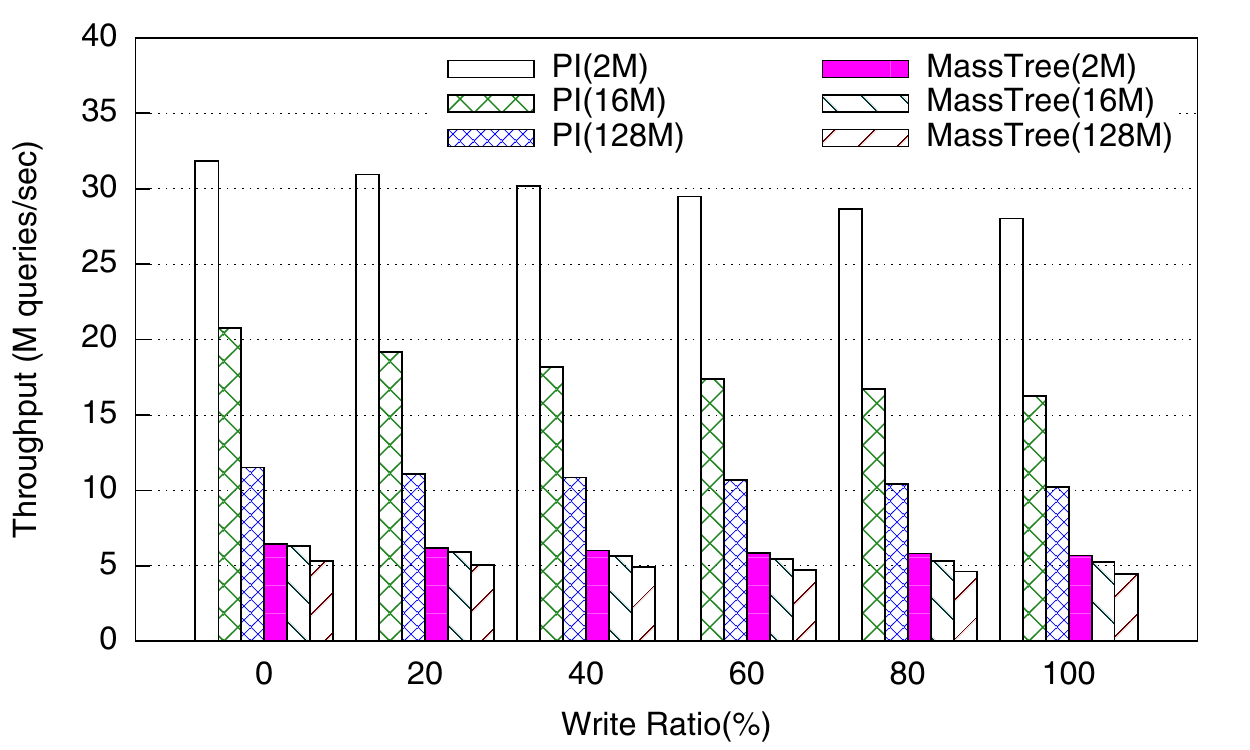}
		\label{fig:ycsb:updateratio0.5}
	}
	\subfigure[$\theta=0.9$] {
		\includegraphics[width=.40\textwidth]{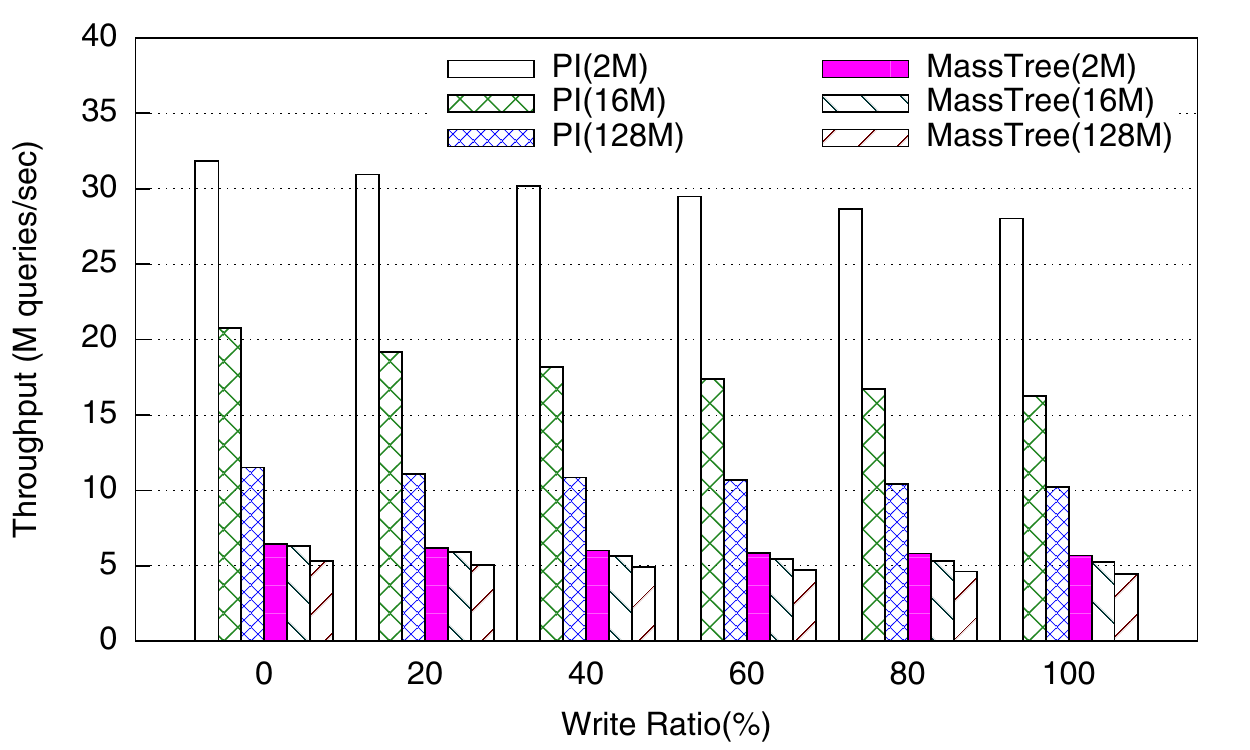}
		\label{fig:ycsb:updateratio0.9}
	}
	\caption{Query throughput vs query skewness (with self-adjusted threading)}
	\label{fig:ycsbratio}
\end{figure*}

\begin{figure*}[tb]
	\centering
	\subfigure[$\theta=0.5$] {
		\includegraphics[width=.40\textwidth]{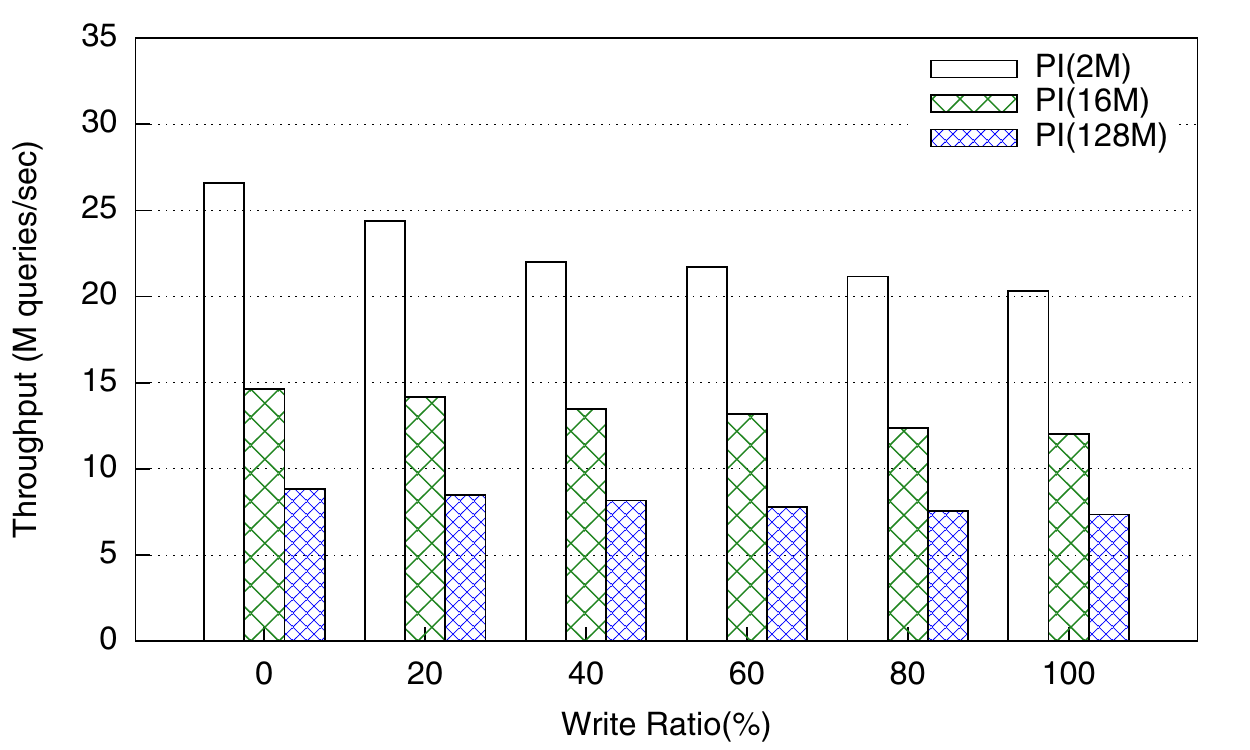}
		\label{fig:ycsb:uniformratio0.5}
	}
	\subfigure[$\theta=0.9$] {
		\includegraphics[width=.40\textwidth]{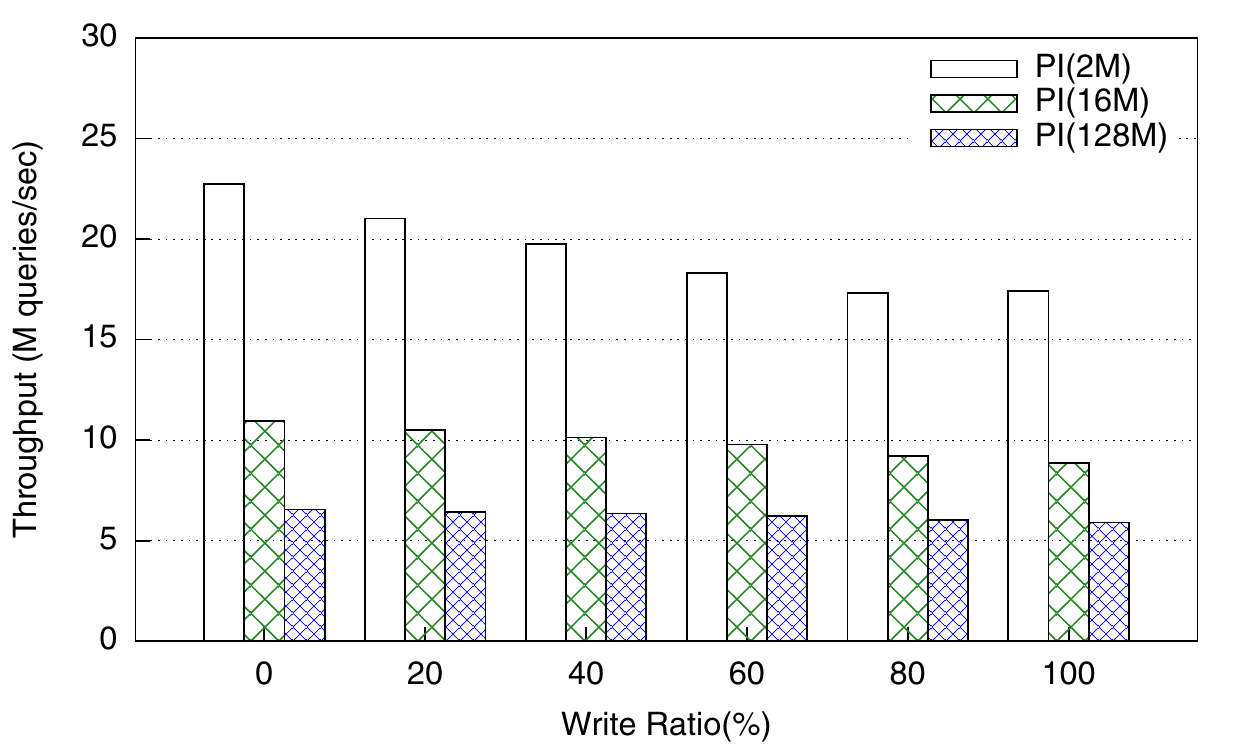}
		\label{fig:ycsb:uniformratio0.9}
	}
	\caption{Query throughput vs query skewness (without self-adjusted threading)}
	\label{fig:threading}
\end{figure*}
%%%% ooibc3: this should have been said
%%%%       we add it to the default experimental settings
In this section, PI's performance of query processing is explored
in the presence of query skewness.
As before, there are eight threads running over
the four NUMA nodes, and three datasets with default sizes
% {\color{blue}
(2M, 16M and 128M)
%}
 are used.
The skewness in query workload 
is realized via varying the probability parameter, $\theta$, of zipfian distribution
for workload generation.
An intuitive impression on the skewness of the query
workloads used in this section is given in Appendix~\ref{sub:ycsbworkload} .%subFigure~\ref{fig:heatmap}.

Figure~\ref{fig:ycsbratio} exhibits the variation in the throughput
query processing with respect to query skewness and update ratio in
query workloads. By comparing this figure with Figure~\ref{fig:mix},
we can observe that query skewness has only a little impact on the 
performance of PI in terms of query processing. 
We attribute this resistance to query skewness of PI to the self-adjusted
threading mechanism presented in Section~\ref{subsec:load}, which 
dynamically allocates computing resources among NUMA nodes 
based on the query load on each node. In fact, for the query workload
with zipfian probability parameter $\theta = 0.5$, the numbers of threads
spawned at four NUMA nodes are 3, 2, 2 and 1, respectively, and for 
the other query workload with $\theta = 0.9$,
these numbers become 4, 2, 1 and 1, respectively.
For comparison purposes, the corresponding result measured with 
the threading mechanism disabled is shown in Figure~\ref{fig:threading}, 
from which one can see that the threading mechanism does significantly 
enhance the throughput.

It should also be noted that the skewness in query workload does not
always exert a negative impact on the throughput.
When fed with a workload consisting of pure search queries against keys
following a zipfian distribution with $\theta = 0.5$, PI is able to achieve
a throughput that is even higher than what is achieved in the case of no query 
skewness, as shown in Figure~\ref{fig:mix} and \ref{fig:ycsb:updateratio0.5}.
This is probably because in
the NUMA node with the most amount of query load, each of the three spawned
thread accesses only
an even restricted portion of the index, and hence can utilize the cache % WWF: what's 'even restricted'?
more efficiently.
 
\subsection{Details on YCSB Workload}
\label{sub:ycsbworkload}

\begin{figure*}[!htbp]
	\centering
	\subfigure[$\theta=0$] {
		\includegraphics[width=.28\textwidth]{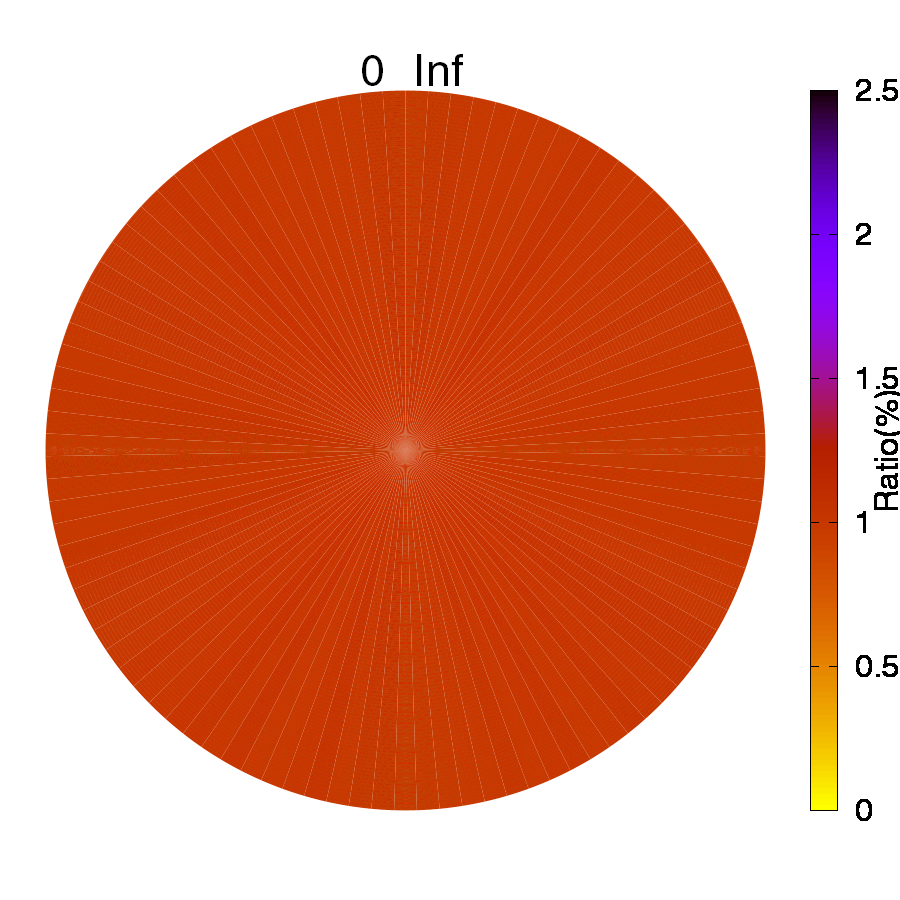}
		\label{fig:heatmap0}
	}
	\subfigure[$\theta=0.5$] {
		\includegraphics[width=.28\textwidth]{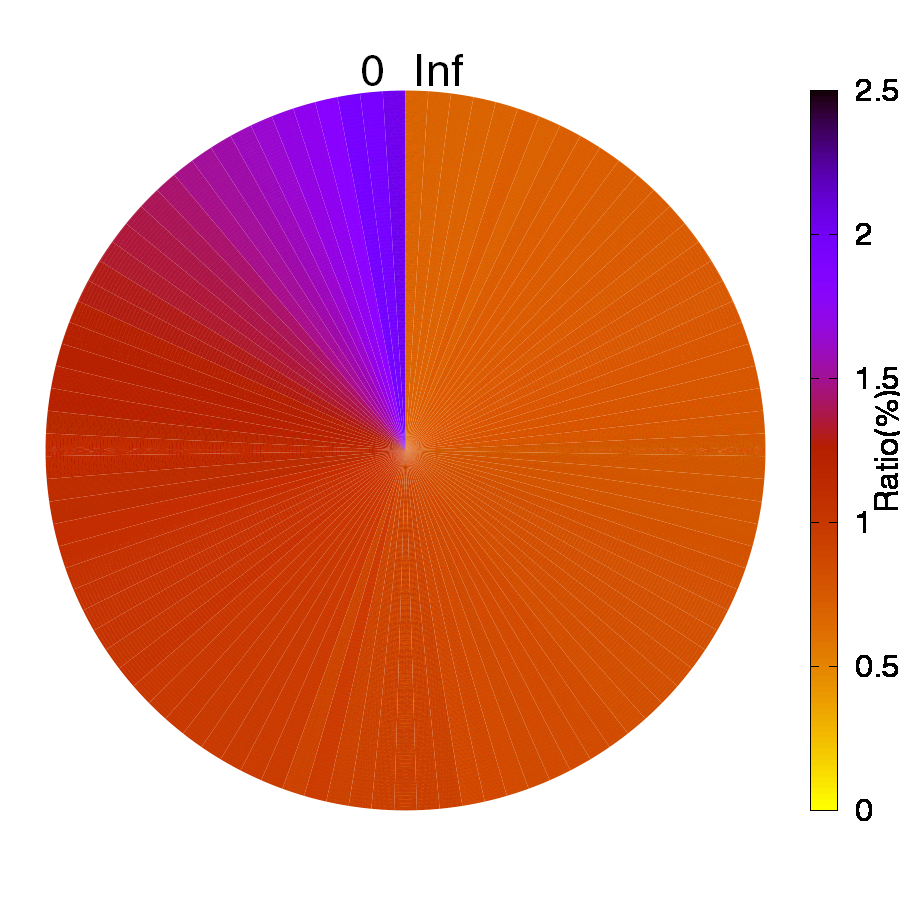}
		\label{fig:heatmap5}
	}
	\subfigure[$\theta=0.9$] {
		\includegraphics[width=.28\textwidth]{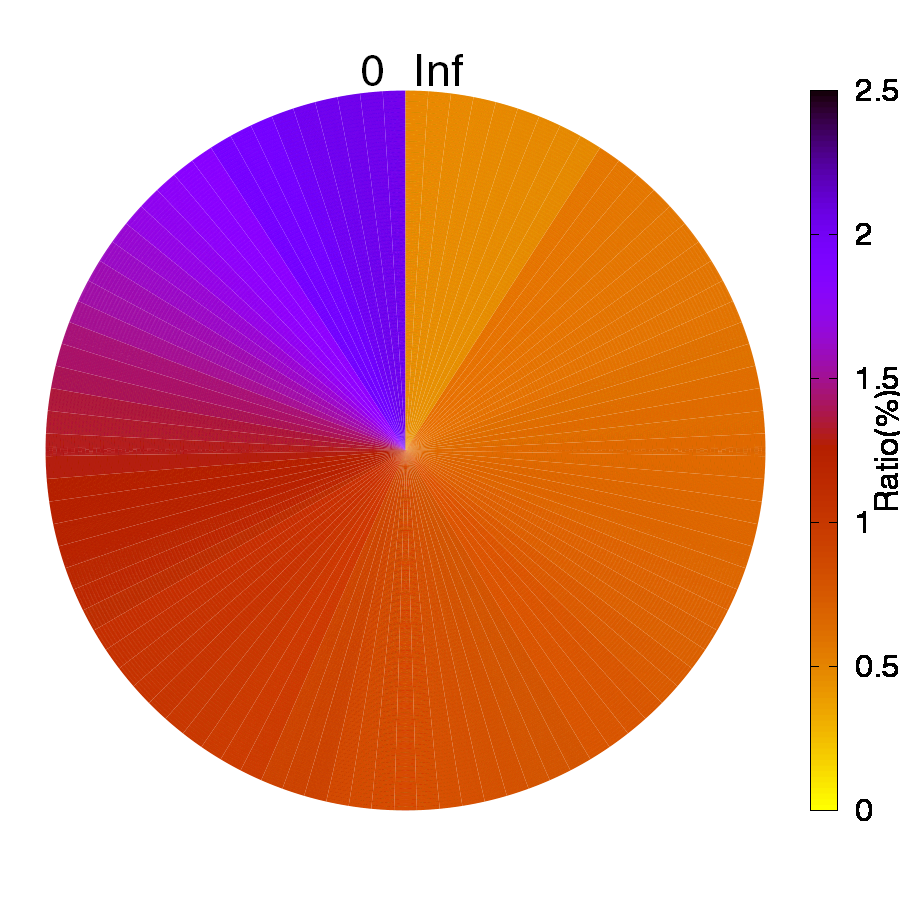}
		\label{fig:heatmap9}
	}
	\caption{Key distribution in three YCSB query workloads}
	\label{fig:heatmap}
\end{figure*}

\begin{figure*}[!htb]
	\centering
	\subfigure[workload A] {
		\includegraphics[width=.3\textwidth]{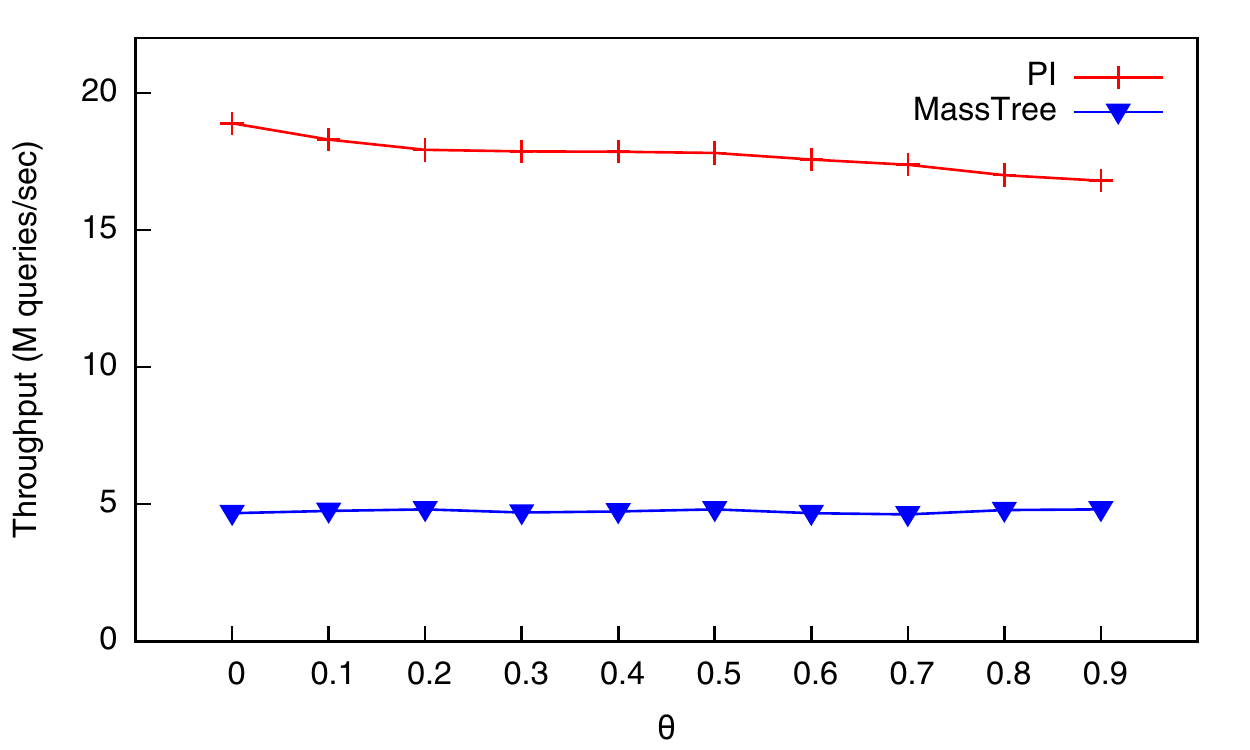}
		\label{fig:workloadA}
	}
	\subfigure[workload B] {
		\includegraphics[width=.3\textwidth]{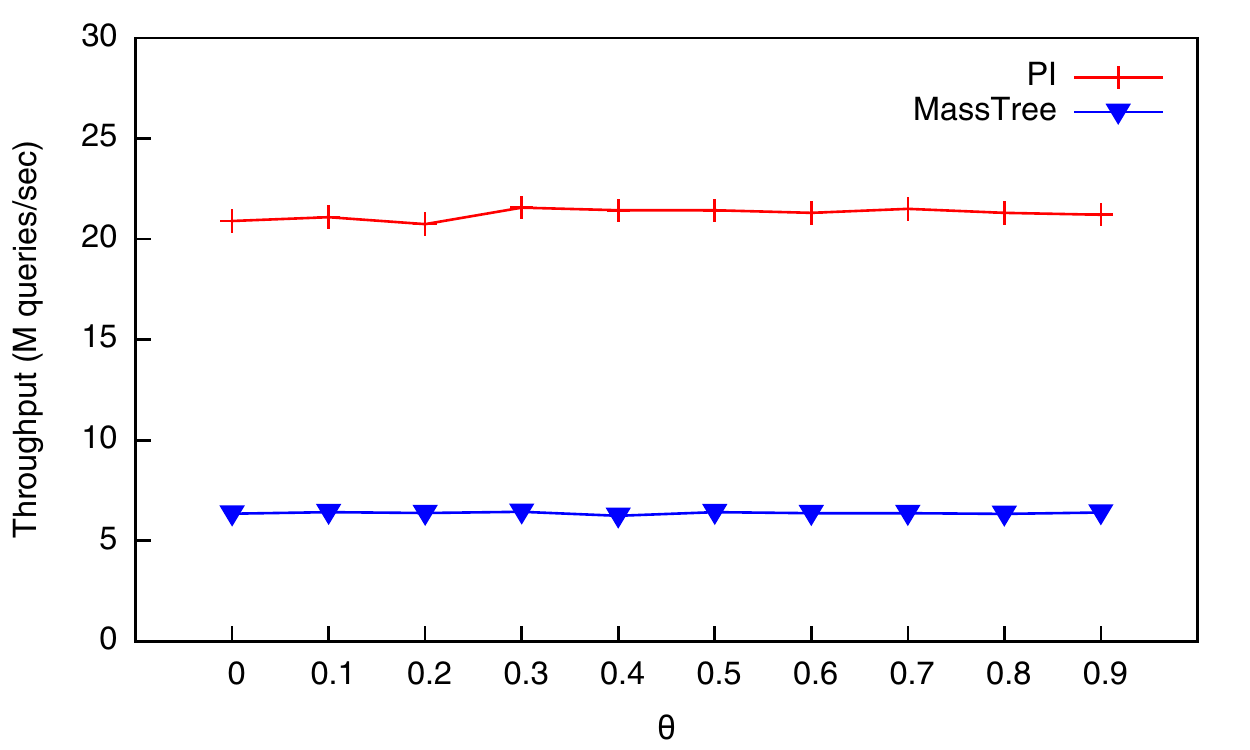}
		\label{fig:workloadB}
	}
	\subfigure[workload C] {
		\includegraphics[width=.3\textwidth]{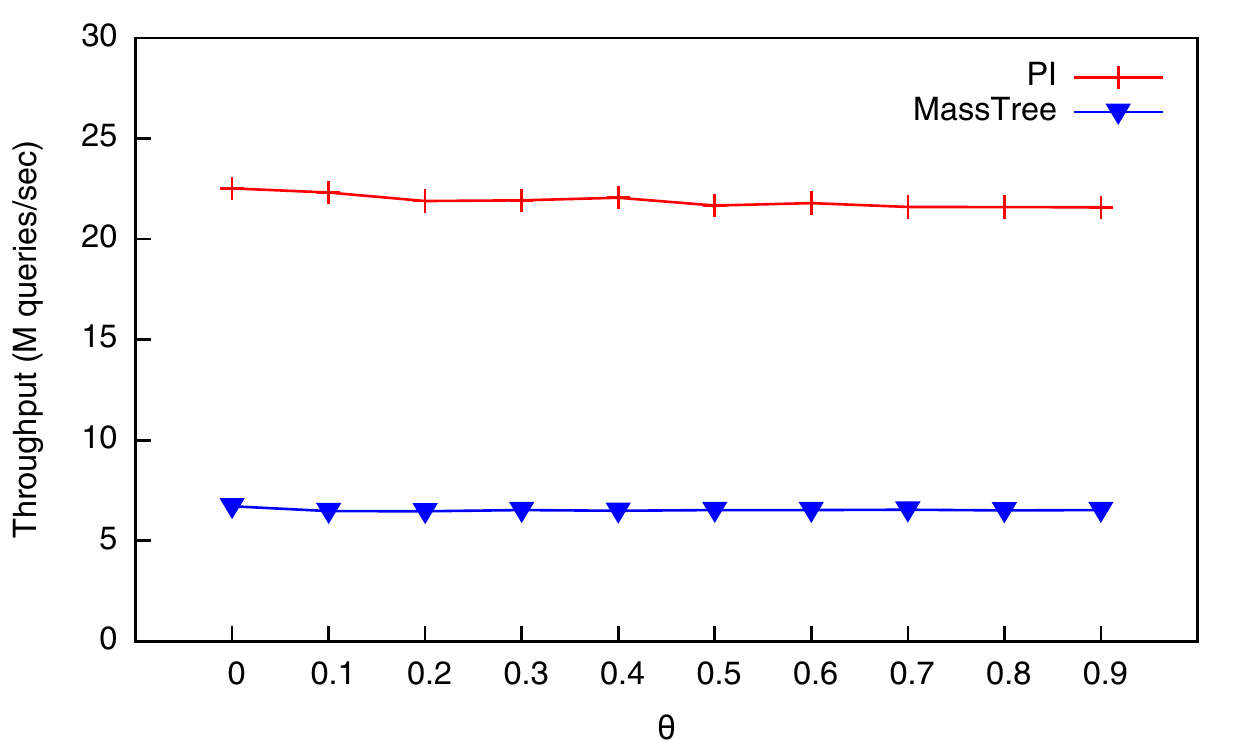}
		\label{fig:workloadC}
	}
	\caption{Query throughput vs skewness}
	\label{fig:workload}
\end{figure*}

Figure~\ref{fig:heatmap} shows the variance of skewness in the key distribution
of the three YCSB workloads for the experiment of
Section~\ref{ssub:skew}. 
%{\color{blue}
In this figure, each circle consists of 100 sectors separating the whole
key space into 100 disjoint ranges with equal coverage, 
and the color of a sector represents how frequently the keys within the
relative range are queried. 
%	}
Since the keys in each YCSB workload follow a zipfian
distribution, the workload becomes more skewed with the increase of probability
parameter $\theta$. However, as shown in Figure~\ref{fig:ycsbratio}, the
skewness in the query workload rarely affects the performance of PI
%{\color{blue}
(as well as Masstree)
%}
, which
is also validated in Figure~\ref{fig:workload}, where the zipfian parameter in
three YCSB query workloads, each with a different update ratio (0.5, 0.05 and 0,
respectively), are varied from 0 to 0.9 
to investigate how PI can adapt to query skewness. 

\subsection{Range Queries}
\label{sub:range}

\begin{figure}[tbp]
	\centering
	\includegraphics[width=0.8\linewidth]{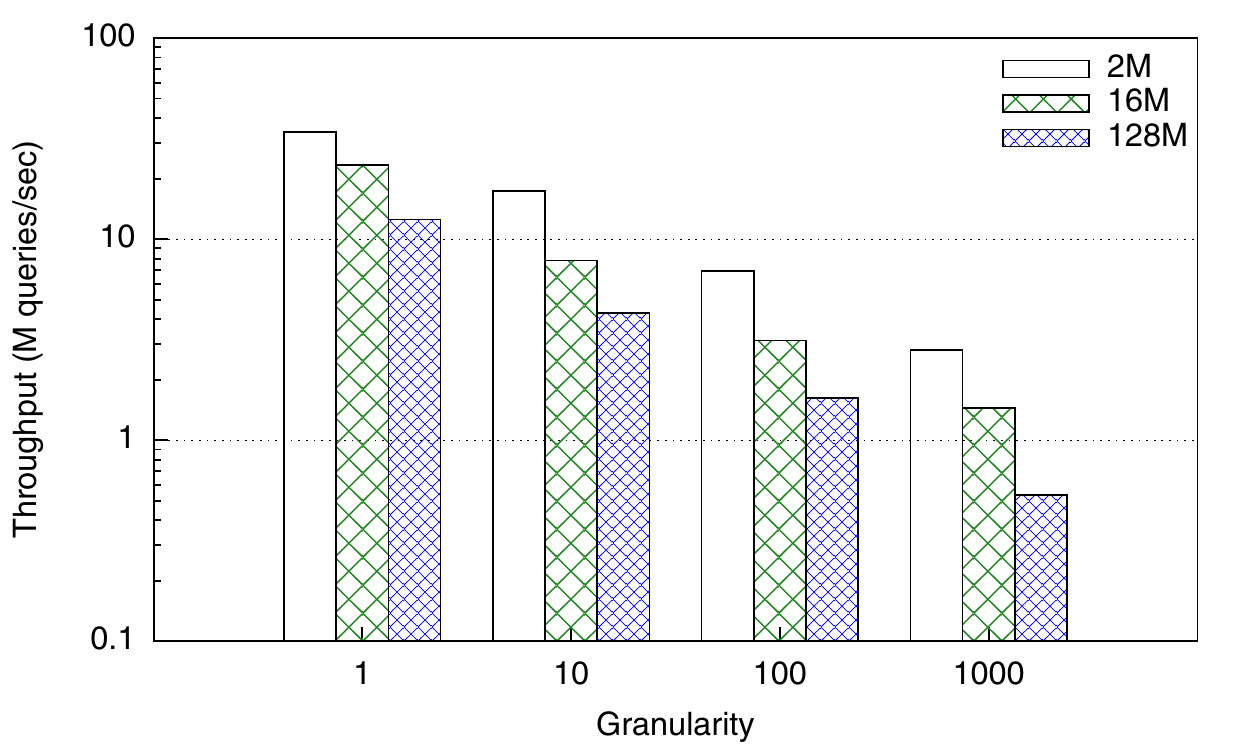}
	\caption{Query throughput of range query}
	\label{fig:range}
\end{figure}
Figure~\ref{fig:range} describes how the performance of range queries
varies with granularity, by which we mean the average number of results (data nodes)
returned for a given range query. In this experiment, only the performance of
search queries is explored, and the same number of query batches, each with 8192
range queries, are issued for each dataset.
For comparison, the result for point query (granularity = 1)
is also shown in this figure.

%%% ooibc: what are the effects of selectivity?

It can be observed from Figure~\ref{fig:range} that query throughput decreases
with the granularity of range query at a constant rate for each dataset.
Due to the better cache utilization of the index built from smaller datasets, the
query throughput decreases a little more slowly for smaller datasets than
for larger datasets. For the granularity of 1000, 
%%% ooibc: 1000 of what? results? out of?
the processing of each query batch
accesses almost 8M of data nodes, which means there are many data nodes being accessed
multiple times for the small and medium datasets with 2M and 16M keys, respectively.
Hence, the number of slow memory accesses incurred by reading entries and data nodes
is further reduced by a larger amount for these two datasets than for the large dataset,
and this is the reason to that the search throughput for the large dataset drops much
faster than that for the other two datasets when the granularity varies from 100 to 1000.

\subsection{Effects of Optimizations}
\label{sub:effect_of_different_optimizations}

We now explore how the optimization techniques such as
SIMD processing, NUMA-aware index partitioning,
prefetching and group query processing,
affect the performance of PI.
The impact of organizing queries into batches
has already been studied in Section~\ref{subsec:batch},
and hence is not discussed here. The dataset used for this
experiment contains 16M keys, and the other parameters
are set to the default sizes.

\begin{figure}[tbp]
	\centering
	\includegraphics[width=0.8\linewidth]{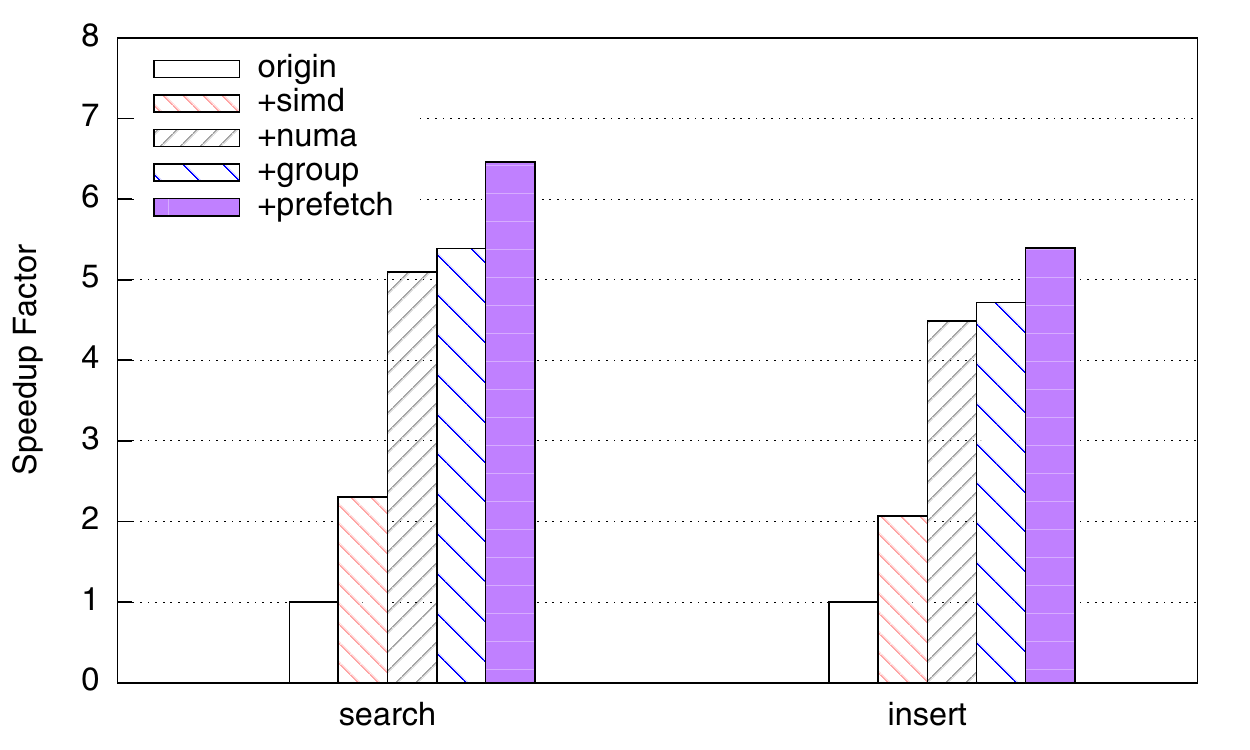}
	\caption{Effects of optimizations}
	\label{fig:speedup}
\end{figure}

Figure~\ref{fig:speedup} shows the breakdown of the gap
between the query performance of PI and a typical
skip list with none of the optimizations enabled.
By grouping the keys appearing at higher layers into entries,
and leveraging SIMD to significantly reduce the number
of key comparisons and hence memory/cache accesses,
the throughput of PI can be improved by 1.3x and 1x
for search and insert queries, respectively.
NUMA-aware operation leads to another huge performance gain,
improving the query throughput by 1.2x for both kinds
of queries. This performance gain is because our NUMA-aware
optimization largely eliminates
accessing the memory of remote NUMA nodes,
which is several times slower than accessing local memory.
Group query processing brings in a slight improvement of 0.05x
in query throughput, and prefetching contributes to the final
performance gain of 0.2x and 0.14x in the throughput of search and insert 
queries, respectively.

\section{conclusion and Future Work}
\label{sec::conclusion}

In this paper, we argue that skip list, due to its high parallezability, is a better candidate for in-memory index
than $B^+$-tree in concurrent environment. Based on this argument, we propose
PI, a cache-friendly, latch-free index that supports both point query and
range query. PI consists of an index layer, which is in charge of key search,
and a storage layer responsible for data retrieval, and the layout of the index
layer is carefully designed such that SIMD processing can be applied to 
accelerate key search.  The experimental results show that PI is three times faster than
Masstree in terms of query throughput. 

For future work, we seek to implement a finer-grained mechanism for the
rebuilding of the index layer, which is currently conducted against the whole
index layer and thus not necessary in the presence of skewed queries which only
update a small portion of the index layer. In addition, we are also exploring
applying several other optimizations to PI. For instance, cache locality can be
further enhanced by pinning the high levels of the
index layer in the cache to prevent them from being evicted in the case of
insufficient cache space, and a large memory page size can reduce the number of
TLB misses incurred by memory accesses.
%\section{Acknowledgments}

\bibliographystyle{abbrv}
\bibliography{sigproc}  

\end{document}